\documentclass[aps,prb,twocolumn,amsfonts,superscriptaddress]{revtex4-1}

\usepackage[latin1]{inputenc}
\usepackage{amsmath,amssymb}
\usepackage{graphicx,color}
\usepackage{braket}
\usepackage{subfigure}
\usepackage{bm}

\usepackage{titlesec}
\titleformat{\section}[runin]{\normalfont\itshape}{}{3pt}{}[.]

\usepackage[breaklinks]{hyperref}
\usepackage[dvipsnames]{xcolor}
\hypersetup{colorlinks,citecolor=blue,filecolor=blue,linkcolor=blue,urlcolor=blue}

\usepackage{natbib}
\setcitestyle{square,numbers}

\newcommand{\eg}[0]{\textit{e.g. }}

\newcommand{\hc}[0]{\textrm{H.c.}}
\newcommand{\mf}[1]{\mathfrak{#1}}
\newcommand{\mc}[1]{\mathcal{#1}}
\newcommand{\mb}[1]{\mathbb{#1}}
\newcommand{\trm}[1]{\textrm{#1}}

\newcommand{\tbf}[1]{\textbf{#1}}
\renewcommand{\vec}[1]{\bm{#1}}

\graphicspath{{../figs/}}
\allowdisplaybreaks[0]

\begin{document}

\title{Floquet  Majorana Corner  States Driven by Ferromagnetic Resonance}
\title{Floquet  Helical Topological Superconductivity Driven by Ferromagnetic Resonance}
\title{Floquet  Second-Order Topological Superconductor \\ Driven via Ferromagnetic Resonance}

\author{Kirill Plekhanov, Manisha Thakurathi, Daniel Loss, and Jelena Klinovaja} \affiliation{Department of Physics, University
  of Basel, Klingelbergstrasse 82, CH-4056 Basel, Switzerland}
\date{\today}
	
\begin{abstract}
We consider a Floquet triple-layer setup composed of a two-dimensional electron gas with spin-orbit interactions, proximity coupled to an $s$-wave superconductor and to  a ferromagnet driven at resonance. The ferromagnetic layer generates a time-oscillating Zeeman field which competes with the induced superconducting gap and leads to a topological phase transition. The resulting Floquet states support a second-order topological superconducting phase with a pair of localized zero-energy Floquet Majorana corner states. Moreover, the phase diagram comprises a Floquet helical topological superconductor, hosting a  Kramers pair of Majorana edge modes  protected by an effective time-reversal symmetry, as well as a  gapless  Floquet Weyl phase. The topological phases are stable against disorder and parameter variations and are within experimental reach.
\end{abstract}

\maketitle

\section{Introduction}

Over the last decade topological states of
matter~\cite{TIRev_HasanKane2010, TIRev_QiZhang2011,
  TIRev_SatoAndo2017, TIRev_WangZhang2017, TIRev_Wen2017} have
attracted a lot of attention. Recently, particular interest has been
raised by higher-order topological insulators and
superconductors~\cite{HOTI_BenalcazarBernevigHughes2017,HOTI_BenalcazarBernevigHughes2017_2,
  HOTI_SongFangFang2017}, which host topologically protected gapless
modes on their higher-order faces (\eg corners in $d>1$, hinges in
$d>2$, with $d$ being the spatial dimension of the system).  However,
such systems were studied mostly at the static
level~\cite{PengBaoOppen2018, ImhofEtAl2018, GeierEtAl2018,
  SchindlerEtAl2018, HsuStanoKlinovajaLoss2018,
  VolpezLossKlinovaja2018,roy1,roy2,Liu2018,Ezawa2018b,kat}, with the
out-of-equilibrium dynamics~\cite{FHOTI_FrancaBrinkFulga2018,
  FHOTI_BomantaraZhouPanGong2018, FHOTI_PengRefael2018,
  FHOTI_HuangLiu2018, FHOTI_RodriguezVegaKumarSeradjeh2018,
  FHOTI_SeshadriDuttaSen2019, FHOTI_NagJuricicRoy2019} rarely
addressed. At the same time, Floquet engineering~\cite{F_Shirley1968,
  F_GoldmanDalibard2014, F_BukovEtAl2015, F_EckardtAnisimivas2015},
based on applying time-periodic perturbations, has been serving as a
powerful tool to generate exotic phases of quantum matter, including
topological and Chern insulators~\cite{FTI_OkaAoki2009,
  FTI_InoueAkihiro2010, FTI_KitagawaEtAl2011,
  FTI_LinderRefaelGalitski2011, FTI_LindnerEtAl2013,
  FTI_CayssolEtAl2013, FTI_PiskunowEtAl2014, KlinovajaStanoLoss2016,
  FTI_KitagawaEtAl2012, FTI_RechtsmanEtAl2013,
  FTI_AidelsburgerEtAl2013, FTI_JotzuEtAl2014, FTI_leHurEtAl2016,
  FATI_KitagawaEtAl2010, FATI_RudnerLidnerBergLevin2013,
  FATI_RoyHarper2017, FATI_MaczewskyEtAl2017, FATI_YaoYanWang2017,
  FATI_GrafTauber2018, FTI_GrushinEtAl2014}, as well as non-Abelian
states such as Majorana bound states (MBSs)~\cite{FTI_JiangEtAl2011,
  FTI_ReynosoFrustaglia2013, FTI_ThakurathiPatelSenDutta2013,
  FTI_LiuLevchenkoBaranger2013, FTI_KunduSeradjeh2013,
  FTI_ThakurathiEtAl2014, FTI_ThakurathiLossKlinovaja2017,
  FTI_KennesEtAl2018,adi}.  Such Floquet phases can be generated by
time-dependent electromagnetic fields~\cite{FTI_OkaAoki2009,
  FTI_InoueAkihiro2010, FTI_KitagawaEtAl2011,
  FTI_LinderRefaelGalitski2011, FTI_LindnerEtAl2013,
  FTI_CayssolEtAl2013, FTI_PiskunowEtAl2014, KlinovajaStanoLoss2016,
  FTI_ThakurathiLossKlinovaja2017, FTI_KennesEtAl2018}. Strong
oscillating electric fields are easy to obtain for this purpose, but
the direct coupling to the spin degree of freedom is more challenging,
typically achieved only indirectly via strong spin orbit
interaction~\cite{Rashba1960, KloeffelLoss2013}.  Hence, finding ways
to generate strong oscillating magnetic fields which couple directly
to the spin is important. One possible solution consists in using
magnetic proximity effects, which, for the static case, have already
been widely studied for superconductors~\cite{FMSC_Buzdin2005,
  FMSC_MooderaEtAl2013} and topological
insulators~\cite{FMTI_KandalaEtAl2013, FMTI_MooderaEtAl2016,
  FMTI_Jiang2016}.

In this work we consider a driven
triple-layer  setup  which
allows us to engineer a Floquet higher-order topological superconductor
(FHOTS). Its key component is a two-dimensional electron gas (2DEG)
with  spin-orbit interaction (SOI) 
sandwiched between
an $s$-wave
superconductor (SC) and a ferromagnet (FM), as shown in
Fig.~\ref{fig:model}(a). The FM layer is resonantly driven by an
external field $\vec{H}(t)$, which results in the generation of an
oscillating magnetic field $\vec{B}(t)$, giving rise to strong Zeeman coupling in the 2DEG. The out-of-plane
component of $\vec{B}(t)$ competes with the proximity-induced
superconducting gap and leads to a topological phase transition in the 2DEG. The
topological phase is identified as Floquet helical topological
superconductor (FHeTS), characterized by the presence of a
Kramers pair of gapless helical Majorana  edge modes protected
by an effective time-reversal symmetry. A moderate in-plane component
of $\vec{B}(t)$ opens a gap in the helical edge modes
through a mass term that changes sign at the corners in a
rectangular geometry, resulting in 
Majorana corner states (MCSs) -- a distinct feature of the
FHOTS. In the regime where the in-plane component of $\vec{B}(t)$ is
strong, the system moves into a gapless Weyl phase. The topological
phase diagram is robust against moderate local disorder, 
detuning from resonance, and  static magnetic fields.

\section{\label{sec:modelB}Model}

\begin{figure}[t]
  \centering
  \includegraphics[width=.99\columnwidth]{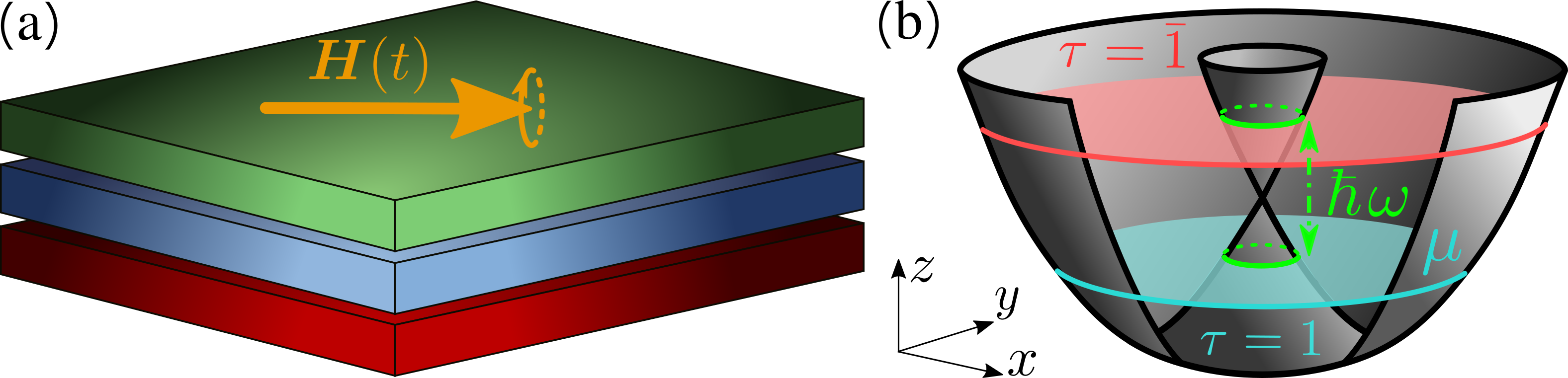}
  \caption{(a) Schematics of the Floquet triple-layer setup consisting of a 2DEG (blue layer) with SOI  of strengths $\alpha_x$ and $\alpha_y$,  proximity coupled to an $s$-wave SC (red layer) at the  bottom and a FM at the top (green layer) which is resonantly
    driven by an external time-dependent magnetic field $\vec{H}(t)$ to generate an  oscillating magnetic field  $\vec{B}(t) = B_{\perp} \cos(\omega t) \vec{e}_z + B_{\parallel}
    \sin(\omega t) \vec{u}_{\parallel}$ in the 2DEG. (b) Band structure of the 2DEG in the isotropic regime $\alpha_x = \alpha_y$. The chemical potential $\mu$ is fixed below the crossing point of the spin-split bands (indicated by the blue area). The driving frequency $\omega$ is tuned to achieve resonance at the smallest Fermi momentum (represented by the two green  circles) between the two Floquet bands labeled by  $\tau=\pm 1$.}
  \label{fig:model}
\end{figure}

The triple-layer setup is composed of a 2DEG with strong Rashba and
Dresselhaus SOIs, proximity coupled to an $s$-wave SC and a FM, as
shown in Fig.~\ref{fig:model}(a). We assume that the SOI vector is
along $z$ direction, which is perpendicular to the  $xy$ 2DEG-plane. The SOI
coupling strengths $\alpha_x = \alpha_R + \beta_D$ and
$ \alpha_y = \alpha_R - \beta_D$ are expressed in terms of the Rashba
and Dresselhaus SOI coefficients $\alpha_R$ and $\beta_D$ for a proper
choice of the coordinate system~\cite{NittaAkazakiTakayanagiEnoki1997,
  EngelsLangeSchapersLuth1997, SchliemannEguesLoss2003, Winkler2003,
  BernevigOrensteinZhang2006, KoralekEtAl2009, DuckheimMaslovLoss2009,
  MengKlinovajaLoss2014}. 
 Introducing a creation operator $\psi^{\dag}_{\sigma \vec{k}}$  acting on
an electron with momentum $\vec{k}=(k_x, k_y)$ and spin component $\sigma$
along the $z$ axis, the corresponding Hamiltonian reads
\begin{align}
  \label{eq:hamSocRw}
 &H_{0} =
  \sum\limits_{\sigma\sigma'}
  \int \trm{d} \vec{k}\
  \psi^{\dag}_{\sigma \vec{k}}
  \bigg[ \frac{\hbar^2 \vec{k}^2}{2 m} - \mu \\ 
  &\hspace{115pt}-
    \alpha_x k_x \sigma_y
   + \alpha_y k_y \sigma_x
    \bigg]_{\sigma\sigma'}\!\!\!
  \psi_{\sigma' \vec{k}}
  \;. \nonumber
\end{align}
Here,  $\sigma_{j}$ are the Pauli matrices acting in
spin space. The chemical potential $\mu$ is calculated from the
crossing point of the two spin-split bands at ${\vec k}=0$. In the following, it will be convenient to introduce the SOI energy $E_{\trm{so}} = \hbar^2 k_{\trm{so}}^2 / (2 m)$ and the SOI
momentum $k_{\trm{so}} = m \alpha_x / \hbar^2$. We note that this Hamiltonian effectively describes a 2D array of coupled Rashba wires
if the mass $m$ is also chosen to be anistropic in the $xy$ plane such
that $m_y \neq m_x$ ~\cite{PoilblancEtAl1987,
  KaneMukhopadhyayLubensky2002, GorkovLebed1995, KlinovajaLoss2013,
  JeffreyKane2014, KlinovajaLoss2014, MengStanoKlinovajaLoss2014,
  NeupertChaminMudryThomale2014, SagiOreg2014,
  KlinovajaTserkovnyakLoss2015, KlinovajaStanoLoss2016,
  SantosEtAl2015, HuangEtAl2016}.

The proximity effect between the 2DEG and the SC is described by the
following Hamiltonian
\begin{equation}
  H_{\trm{sc}} = \frac{\Delta_{\trm{sc}}}{2}
  \sum\limits_{\sigma\sigma'} \int \trm{d} \vec{k}
  \left(
    \psi^{\dag}_{\sigma \vec{k}}
    \left[ i \sigma_y \right]_{\sigma\sigma'}
    \psi^{\dag}_{\sigma'(-\vec{k})} + \hc
  \right),
\end{equation}
where $\Delta_{\trm{sc}}$ is the induced SC gap. The resulting 2DEG-SC
heterostructure is placed in the vicinity of a FM layer, and the setup
is subjected to an external magnetic field $\vec{H}(t)$. Under the FM resonance
condition (see discussion below), the FM generates an oscillating demagnetizing field which
adds up to $\vec{H}(t)$ to produce a total magnetic field
$\vec{B}(t) = B_{\perp} \cos(\omega t) \vec{e}_z + B_{\parallel}
\sin(\omega t) \vec{u}_{\parallel}$ in the 2DEG. Here,
$\omega = 2 \pi / T$ denotes the oscillation frequency and the 2D
vector $\vec{u}_{\parallel}=(u_x, u_y)$ indicates the orientation of
the magnetic field in the $xy$ plane. The FM proximity effect is
described by the following Floquet-Zeeman term
\begin{align}
  \label{eq:floquetTermB}
  H_{\trm{Z}}(t) =
  2 \sum\limits_{\sigma\sigma'} \int \trm{d} \vec{k}\
  &
    \psi^{\dag}_{\sigma \vec{k}}
    \Big( t_{\trm{Z}}^{\perp} \cos(\omega t) [\sigma_z]_{\sigma\sigma'}
    \notag \\ +
  & t_{\trm{Z}}^{\parallel}
    \sin(\omega t) [ \vec{u}_{\parallel} \cdot \vec{\sigma} ]_{\sigma\sigma'}
    \Big)  \psi_{\sigma' \vec{k}},
\end{align}
where ${t_{\trm{Z}}^{\nu} = \mu_B g_{\nu} B_{\nu} / 2}$ (with
$\nu=\parallel, \perp$) are two Floquet-Zeeman amplitudes. The
anisotropy in the $g$-factors, which leads to $g_{\parallel}$ and
$g_{\perp}$, arises from the quantum confinement and the intrinsic
strain of the 2DEG~\cite{LeJeuneEtAl1997, MalinowskiHarley2000,
  TolozaEtAl2012}.

The resulting time-dependent problem can be solved using the Floquet
formalism~\cite{F_Shirley1968, F_GoldmanDalibard2014, F_BukovEtAl2015,
  F_EckardtAnisimivas2015}, by writing the quasi-energy operator
${H = H_{0}+H_{\trm{sc}}+H_{\trm{Z}}(t) - i\hbar\partial_t}$ in
the Floquet-Hilbert space generated by $T$-periodic states
${\psi_{l \sigma \vec{k}} = \trm{exp}(-i l \omega t)\psi_{\sigma
    \vec{k}}}$, $l \in \mb{Z}$. In this basis $H$ acquires a simple
block-diagonal form, where each block, also referred to as a Floquet
band, is composed of the modes with the same index $l$. The static
term acts within the same block and receives an additional constant
energy shift $\hbar \omega l$, while the oscillating term couples
different blocks. We assume that the chemical potential is restricted
to $-E_{\trm{so}} < \mu < 0$, so that the frequency $\omega$ can be
resonantly tuned to the energy difference between the two spin-split
bands. This allows us to treat the oscillating terms at low energies by
only taking into account the coupling between the modes at $l=0$, to
which we associate a Floquet band index $\tau = 1$, and the modes at
$l=-1$ with $\tau = \bar{1}$.  As a result, in the Nambu basis
$\Psi^{\dag}_{\vec{k}}$ = ($\psi^{\dag}_{1 \uparrow \vec{k}}$, $\psi^{\dag}_{1 \downarrow \vec{k}}$, $\psi^{\dag}_{\bar{1} \uparrow \vec{k}}$, $\psi^{\dag}_{\bar{1} \downarrow \vec{k}}$,
$\psi_{1 \uparrow -\vec{k}}$, $\psi_{1 \downarrow -\vec{k}}$, $\psi_{\bar{1} \uparrow -\vec{k}}$,
$\psi_{\bar{1} \downarrow -\vec{k}}$), the total Hamiltonian reads as $H = \int \trm{d} \vec{k}\ \Psi^{\dag}_{\vec{k}} \mc{H}_{\vec{k}}
\Psi_{\vec{k}} / 2$, with the Hamiltonian density
\begin{align}
  \label{eq:hamDensity}
  & \mc{H}_{\vec{k}} =
    \left[ \frac{\hbar^2 \vec{k}^2}{2m} -
    \mu +  \frac{\hbar \omega (\tau_z - \tau_0)}{2}   \right] \eta_z +
    \Delta_{\trm{sc}} \eta_y \sigma_y +
    \alpha_y k_y \sigma_x   \notag \\
    &
   - \alpha_x k_x \eta_z \sigma_y      +
    t_{\trm{Z}}^{\perp} \eta_z \tau_x \sigma_z +
    t_{\trm{Z}}^{\parallel} \left(
    u_x \tau_y \sigma_x + u_y \eta_z \tau_y \sigma_y \right)
\end{align}
with the Pauli matrices $\tau_i$ ($\eta_i$) acting in the Floquet
(particle-hole) space. In the following, we analyze different
topological phases of the system as a function of the parameters appearing
in Eq.~\eqref{eq:hamDensity}.

\section{Floquet Helical Topological Superconductor}

\begin{figure}[t]
  \centering
  \includegraphics[width=.99\columnwidth]{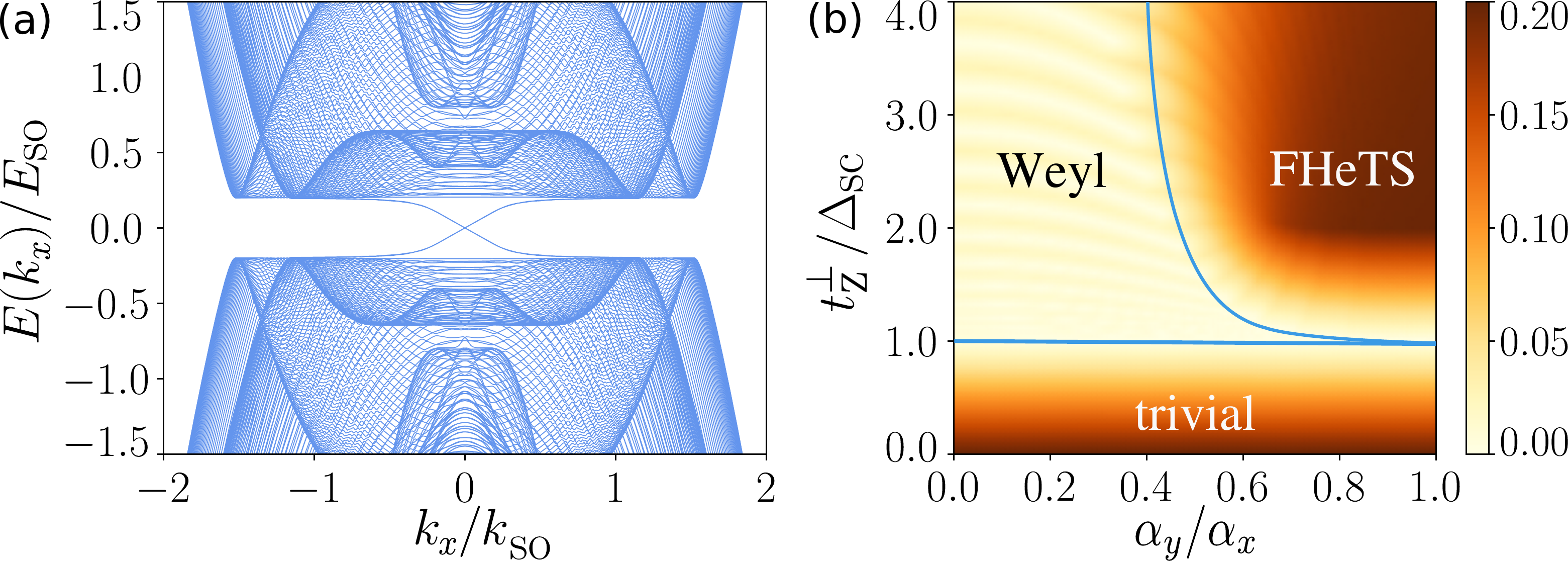}
  \caption{ (a) Energy spectrum $E(k_x)$ in the FHeTS phase with
    $t_{\trm{Z}}^{\perp} / \Delta_{\trm{sc}} = 3$ and
    $\alpha_x = \alpha_y$. The helical edge modes are localized at the
    edges. {(b)} Phase diagram showing the gap to the first excited
    bulk state (in units of $E_{\trm{so}}$) as a function of the
    ratios $\alpha_y / \alpha_x$ and
    $t_{\trm{Z}}^{\perp} / \Delta_{\trm{sc}}$. Blue lines indicate the
    phase boundaries. In the isotropic regime, $\alpha_x = \alpha_y$,
    the phase transition occurs at the critical point
    $\Delta_{\trm{sc}} = t_{\trm{Z}}^{\perp}$. When
    $\alpha_x \neq \alpha_y$, the critical point is transformed into a
    gapless Weyl phase, and the value
    $t_{\trm{Z}}^{\perp} / \Delta_{\trm{sc}}$ required to reach the
    FHeTS increases up to the point of strong anisotropy beyond which
    the FHeTS cannot be reached. Remaining parameters in both
    simulations are $t_{\trm{Z}}^{\parallel} = 0$,
    $\Delta_{\trm{sc}} / E_{\trm{so}} = 0.2$,
    $k_{\trm{so}} L_x = k_{\trm{so}} L_y = 80$, and
    $\mu =-E_{\trm{so}}/2$.}
  \label{fig:ftsc}
\end{figure}

In order to determine the phase diagram of our model, we first
consider the effect of the out-of-plane component of $\vec{B}(t)$ by
imposing $t_{\trm{Z}}^{\parallel} = 0$ in
Eq.~\eqref{eq:hamDensity}. In the isotropic regime with
$\alpha_x = \alpha_y$, the Fermi surface is composed of two concentric
circles and the problem depends only on the magnitude of the momentum
$|\vec{k}|$, as shown in Fig.~\ref{fig:model}(b). The resonance
condition for the frequency $\omega$ is satisfied along the entire
circle with the smallest Fermi momentum (see the Supplemental Material
(SM)~[\onlinecite{SM}] for more details). The Hamiltonian is
linearized close to the Fermi surface and provides the eigenenergies
$E_1^2 = (\hbar v_F \delta k)^2 + \Delta_{\trm{sc}}^2$ and
$E_{2, \pm}^2 = (\hbar v_F \delta k)^2 + (t_{\trm{Z}}^{\perp} \pm
\Delta_{\trm{sc}})^2$, with $v_F = \alpha_x / \hbar$ the Fermi
velocity and $\delta k$  the radial distance from the Fermi
surface. The phase diagram consists of two gapped phases separated by
the gapless line $\Delta_{\trm{sc}} = t_{\trm{Z}}^{\perp}$. The
topologically trivial (topological) phase is identified with the
regime $\Delta_{\trm{sc}} > t_{\trm{Z}}^{\perp}$
($\Delta_{\trm{sc}} < t_{\trm{Z}}^{\perp}$). For
$t_{\trm{Z}}^{\parallel} = 0$, the system is characterized by an
emerging effective time-reversal symmetry
$T_{\trm{eff}} = -i \tau_z \sigma_y \mc{K}$, a particle-hole symmetry
$P = \eta_{x} \mc{K}$, and a chiral symmetry $U_{C} = PT_{\trm{eff}}$,
with  $\mc{K}$ the complex-conjugation operator. Thus, the system
belongs to the DIII symmetry class with $\mb{Z}_2$ topological
invariant.

The topological phase, denoted as FHeTS, hosts gapless boundary modes
 -- a Kramers pair of Floquet Majorana fermions
$\ket{\Phi_{\pm}}$, obeying $P \ket{\Phi_{\pm}} = \ket{\Phi_{\pm}}$
and $T_{\trm{eff}} \ket{\Phi_{\pm}} = \pm \ket{\Phi_{\mp}}$. The Kramers partners
propagate in opposite directions along the same edge, forming a pair of
helical modes protected by the effective time-reversal and
particle-hole symmetries. We have verified the presence of these modes 
numerically in the discretized version of the model (see the SM~[\onlinecite{SM}]) defined
on a rectangular lattice of size $L_x \times L_y$ with periodic
boundary conditions along $x$, 
as shown in
Fig.~\ref{fig:ftsc}(a).

If the rotation symmetry is broken ($\alpha_x \neq \alpha_y$), the
resonance condition can be satisfied only along a particular direction
in  momentum space, resulting in an off-set
$\delta \omega$ in the resonance condition almost everywhere except at a
few points on the Fermi surface. While a small $\delta \omega$ in the
weak anisotropy regime hardly affects the phase diagram and can be
compensated by increasing $t_{\trm{Z}}^{\perp}$, strong anisotropy
effects are more drastic. In Fig.~\ref{fig:ftsc}(b), we calculate
numerically the energy of the lowest bulk state as a function of the
ratios $\alpha_y / \alpha_x$ and
$t_{\trm{Z}}^{\perp} / \Delta_{\trm{sc}}$. We see that the critical
point $\Delta_{\trm{sc}} = t_{\trm{Z}}^{\perp}$ transforms into a
gapless Weyl phase at a finite value of anisotropy
$(\alpha_y -\alpha_x) / \alpha_x$. This gapless regime is
characterized by a semi-metal energy structure with four Weyl
cones. The nodes of the Weyl cones appear first on the Fermi surface
and move further in the reciprocal space, when the parameters
$t_{\trm{Z}}^{\perp} / \Delta_{\trm{sc}}$ and $\alpha_y / \alpha_x$
are modified.
We also note that if we decrease the ratio
$t_{\trm{Z}}^{\perp} / \Delta_{\trm{sc}}$ up to a point of reaching
the phase transition to the trivial phase, the low
energy physics becomes insensitive to the anisotropy.

\section{Floquet Majorana corner states}
Next, we analyze the effect of an oscillating
in-plane magnetic field that breaks the effective time-reversal
symmetry $T_{\trm{eff}}$ and, thus, gaps out the helical edge modes of
the FHeTS. Nevertheless, the system
remains topologically non-trivial as it now hosts a set of zero-energy MCSs,
characteristic for the FHOTS phase \cite{fot1}. The presence of such MCSs is uncovered by
focussing on the low-energy degrees of freedom expressed in terms of
the Majorana edge modes $\ket{\Phi_{\pm}}$. The in-plane
Zeeman field $B_{\parallel}$ couples the two helical modes, leading to the following
low-energy Hamiltonian density:
\begin{equation}
  \mc{H}_{\trm{edge}} = \hbar v^{e}_{F} |\vec{k}| \rho_z +
  \tilde{t}^{\parallel}_{\trm{Z}} \rho_y
  \;,
\end{equation}
where the Pauli matrices $\rho_i$ act in the space of $\ket{\Phi_{\pm}}$, $v^{e}_{F}$ is
the velocity of the helical edge modes, and
${\tilde{t}^{\parallel}_{\trm{Z}} = t^{\parallel}_{\trm{Z}} \mf{Im}
  \braket{\Phi_{-} | u_x \tau_y \sigma_x + u_y \eta_z \tau_y \sigma_y
    | \Phi_{+}}}$ is the `mass term'. Thus, our system  is described by
the well-known Jackiw-Rebbi model~\cite{Jackiw1Rebbi976,
  JackiwSchrieffer1981}.

\begin{figure}[t]
  \centering
  \includegraphics[width=.99\columnwidth]{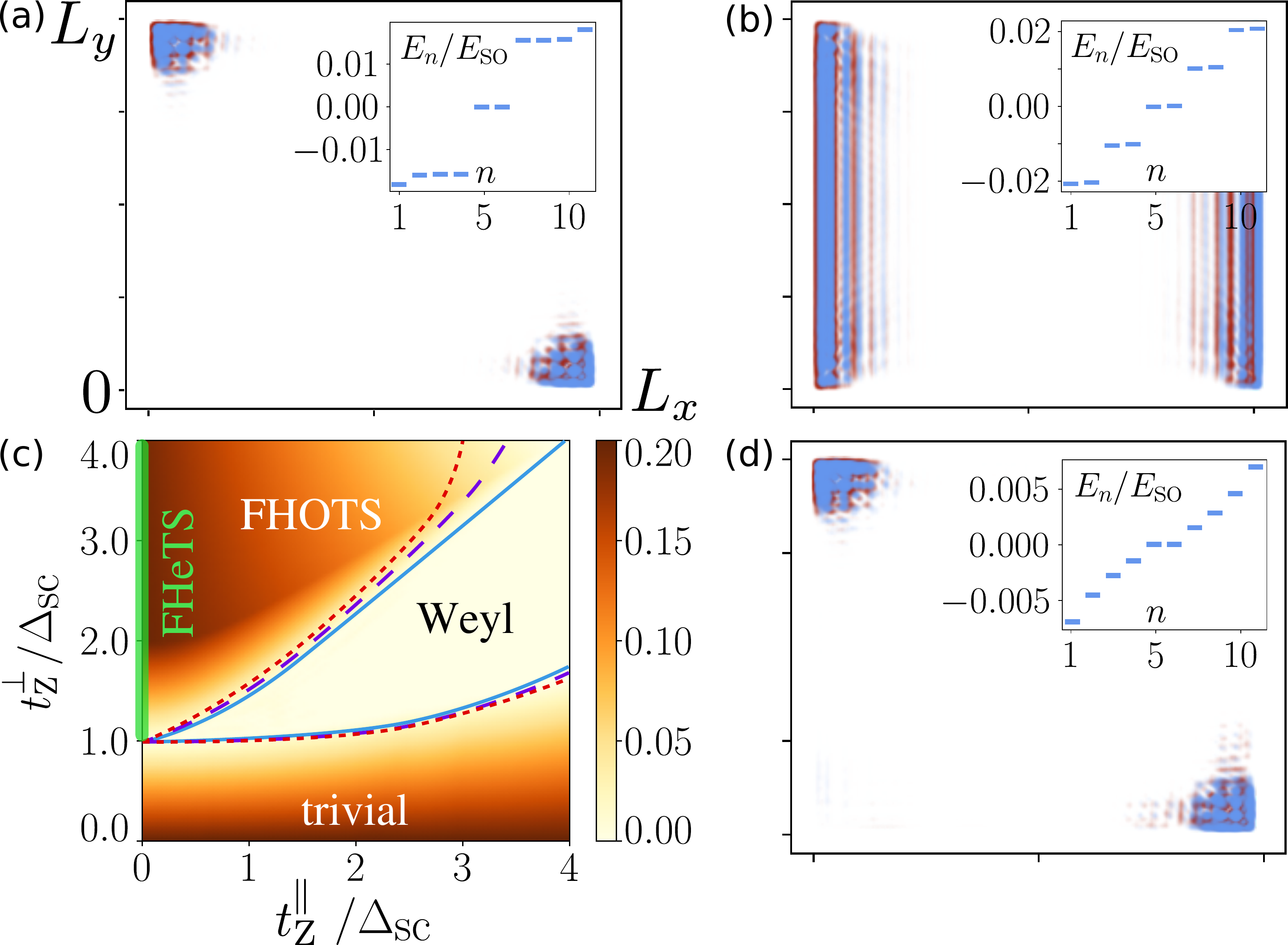}
  \caption{{(a,b,d)} Probability density of the lowest energy state in
    the FHOTS phase for $t^{\perp}_{\trm{Z}} / \Delta_{\trm{sc}} = 3$,
    $t^{\parallel}_{\trm{Z}} / \Delta_{\trm{sc}} = 2.5$, and
    $\alpha_x = \alpha_y$. The inset shows the 10 lowest eigenenergies.
    {(a)} For ${\vec{u}_{\parallel} = (1,1)/\sqrt{2}}$, the in-plane Zeeman
    field $B_{\parallel}$ opens gaps in all edge modes such that zero-energy MCSs emerge at two
    opposite corners. {(b)} The vector $\vec{u}_{\parallel} = (0,1)$ is
    parallel to the edges along which the system stays
    gapless. {(c)} Phase diagram showing the bulk gap (color coded in units of $E_{\trm{so}}$) as a function of the ratios $t^{\perp}_{\trm{Z}} / \Delta_{\trm{sc}}$ and
    $t^{\parallel}_{\trm{Z}} / \Delta_{\trm{sc}}$ with
    ${\vec{u}_{\parallel} = (1,1)/\sqrt{2}}$. The critical point
    $\Delta_{\trm{sc}} = t^{\perp}_{\trm{Z}}$ at
    $t^{\parallel}_{\trm{Z}} = 0$ merges into a gapless Weyl phase
    at finite $t^{\parallel}_{\trm{Z}}$. As a result, higher value of
    $t^{\perp}_{\trm{Z}} / \Delta_{\trm{sc}}$ are required to reach
    the topological phase. 
    The various phase boundaries correspond to 
     $\alpha_y / \alpha_x = 0.8$ (red dotted), $0.9$ (purple dashed),
    and $1.0$ (blue solid).
   {(d)} The  MCSs are stable against moderate
    external perturbations and disorder: 
    $\delta \omega = \sqrt{3} S_{\mu} = \Delta_{\trm{Z}} = 0.10
    E_{\trm{so}}$. Rest of parameters  are the same as in Fig. \ref{fig:ftsc}.
    }
  \label{fig:mcs}
\end{figure}

From the symmetry of the modes $\ket{\Phi_{\pm}}$, we deduce that the
value of the mass term $\tilde{t}^{\parallel}_{\trm{Z}}$ only depends
on the component of the in-plane field $B_{\parallel}$ which is perpendicular
to the corresponding edge in a rectangular geometry (see the
SM~[\onlinecite{SM}]). Generally, the sign of $\tilde{t}^{\parallel}_{\trm{Z}}$ is opposite
on two parallel edges at $x = 0$ ($y=0$) and $x = L_x$ ($y =
L_y$). Hence, $\tilde{t}^{\parallel}_{\trm{Z}}$ has to change its sign at
two opposite corners of the 2DEG, leading to the emergence of domain
walls at these corners that host 
zero-energy 
MCSs, see Fig.~\ref{fig:mcs}(a). In the special case when $\vec{u}_{\parallel}$ is
parallel to one of the edges, the corresponding edge modes stay gapless, see Fig.~\ref{fig:mcs}(b).

The simple boundary description in terms of the Jackiw-Rebbi model is
expected to work in the regime where the amplitude of the in-plane
magnetic field is small. In order to construct the full phase diagram,
we calculated numerically the gap to the first excited bulk state as a
function of the ratios $t^{\perp}_{\trm{Z}} / \Delta_{\trm{sc}}$ and
$t^{\parallel}_{\trm{Z}} / \Delta_{\trm{sc}}$, see
Fig.~\ref{fig:mcs}(c). The FHOTS phase emerges from the FHeTS phase at non-zero
$t^{\parallel}_{\trm{Z}}$. However, if $t^{\parallel}_{\trm{Z}}$ is
large, the system enters into the gapless Weyl phase.
To observe MCSs, the Floquet-Zeeman field perpendicular to the 2DEG
plane should dominate such that the condition
$t_{\trm{Z}}^{\parallel} < t_{\trm{Z}}^{\perp}$ is fulfilled. All the
results remain qualitatively the same for a weak anisotropy
($\alpha_x \neq \alpha_y$), which shifts the topological phase
transition line only slightly. 

\begin{figure}[!t]
  \centering
  \includegraphics[width=.75\columnwidth]{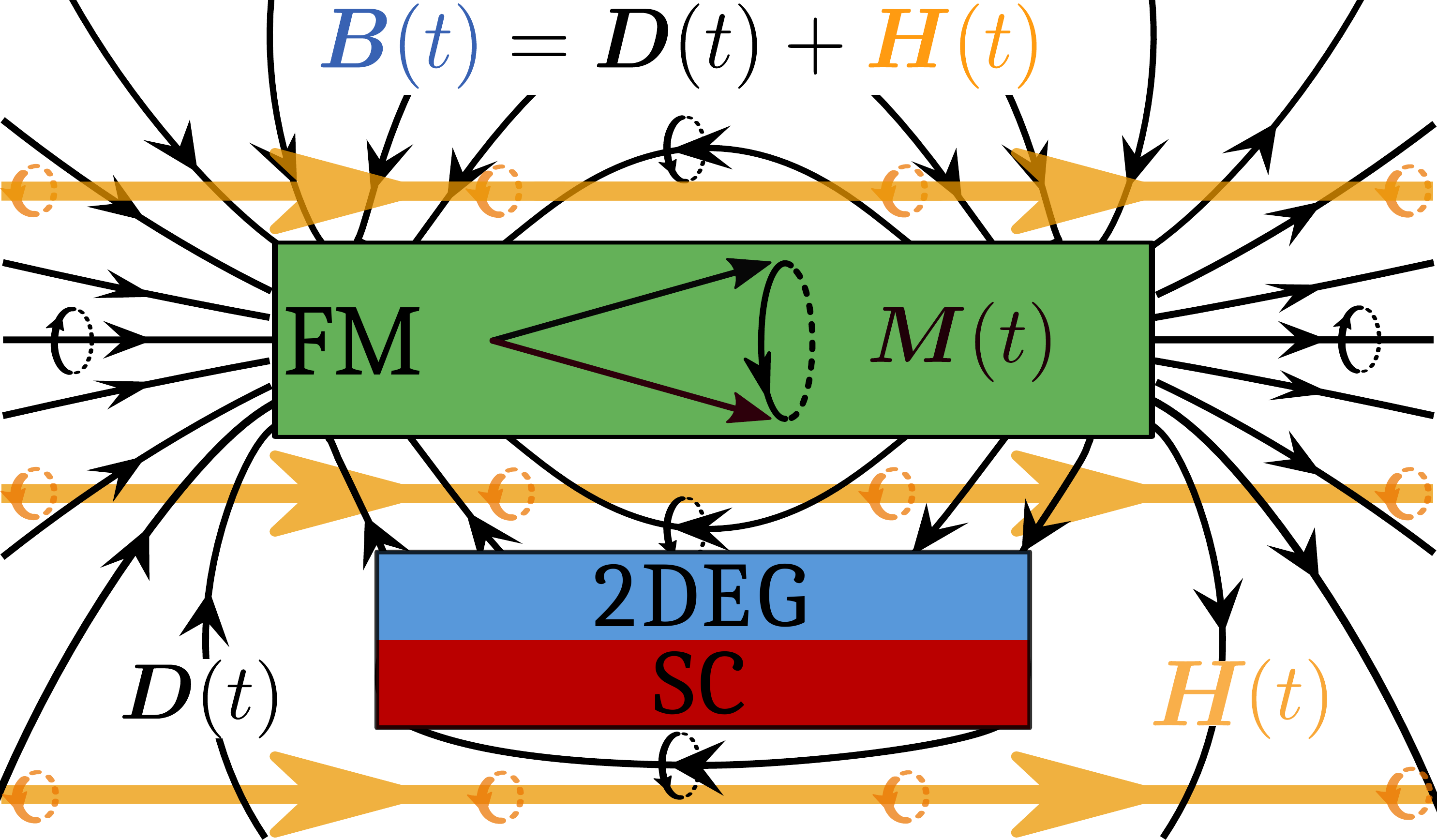}
  \caption{
    Setup of the FM layer to generate the desired
    oscillating magnetic field in the 2DEG. The 2DEG-SC heterostructure is placed
    in the vicinity of the FM. The system is subjected to an external
    magnetic field $\vec{H}(t) = \vec{H}_0 + \vec{h}(t)$ (yellow
    lines) generating FM resonance (shown not to
    scale). This induces the precession of the FM magnetization
    $\vec{M}(t)$ (black lines inside the FM) and demagnetizing field
    $\vec{D}(t)$ (black lines outside of the FM). Close to the surface
    of the FM layer, the static components of $\vec{D}(t)$ and $\vec{H}(t)$
    cancel out, so that the dynamics of the total field
    $\vec{B}(t) = \vec{H}(t) + \vec{D}(t)$ in the 2DEG layer is close to a full
    $360^{\circ}$ rotation.}
  \label{fig:fm}
\end{figure}

\section{Experimental feasibility}

Next, we discuss the stability of our setup. We check numerically that
the topological phases are robust against an off-set $\delta \omega$
in the resonance frequency $\omega$. Similarly, we check the stability
with respect to an on-site disorder by adding a fluctuating chemical
potential randomly chosen from a uniform distribution with  standard
deviation $S_{\mu}$ and with respect to a static magnetic field by
adding a Zeeman term with  strength $\Delta_{\trm{Z}}$ directed
both in-plane and out-of-plane. The result of the calculations is
shown in Fig.~\ref{fig:mcs}(d). The topological phases are
stable against the perturbations of a strength comparable to the
gap. The effect of the out-of-plane component of the
static magnetic field is also less important: the MCSs can be observed
even for $\Delta_{\trm{Z}} \sim t^{\perp}_{\trm{Z}}$ and disappear
only when the static term becomes stronger.

In experiments, the proximity induced SC gap $\Delta_{\trm{sc}}$ is
expected to be of the order of $0.05$~meV, depending on the properties
of the SC and the strength of the coupling between the SC and the
2DEG~\cite{ChangEtAl2015, WanEtAl2015,chris}. As shown in this work,
the strength of the Floquet-Zeeman amplitude $t_{\trm{Z}}^{\perp}$
should exceed $\Delta_{\trm{sc}}$ to reach the topological
regimes. Hence, assuming that the 2DEG material has an electron
$g$-factor $g_{\perp}=15$, the amplitude of the magnetic field
$B_{\perp}$ should be of the order of $0.1$~T. At the same time, the
static component of the magnetic field should be smaller than the
dynamic one and the oscillation frequency $\omega$ should be in the
GHz range.
The FM layer
in the setup is proposed to
generate the required Zeeman fields as follows.
Applying an external magnetic field
$\vec{H}(t) = \vec{H}_0 + \vec{h}(t)$ with
$|\vec{h}(t)| \ll |\vec{H}_0|$ and $\vec{h}(t) \perp \vec{H}_0$ under
 FM resonance condition induces a precession of the FM
magnetization $\vec{M}(t)$~\cite{Kittel1948, Kittel1951}. The
precession cone of $\vec{M}(t)$ depends on the angle between the FM easy axis and
$\vec{H}_0$, while the resonance frequency is determined by the magnitude
$|\vec{H}_0|$. Outside of the FM, the total field $\vec{B}(t)$ is
equal to the sum of the external field $\vec{H}(t)$ and an oscillating
demagnetizing field $\vec{D}(t)$, see
Fig.~\ref{fig:fm}. Hence, by carefully choosing the system geometry,
the static component of $\vec{B}(t)$ could be adjusted close to zero
over a large region of space in the proximity of the FM surface including the 2DEG. The
amplitude of the remaining oscillating component overcomes the
threshold of $0.1$~T inside the 2DEG, as we have confirmed by micromagnetic simulations (see
the SM~[\onlinecite{SM}]). Promising candidates for such FMs are e.g. $\trm{EuS}$~\cite{FMTI_MooderaEtAl2016},
$\trm{GdN}$~\cite{FMTI_KandalaEtAl2013}, and
$\trm{YiG}$~\cite{FMTI_Jiang2016}.

Alternatively, the fast switching or the sustained oscillation of the
FM magnetization has already been achieved experimentally by shining
 optical light on a FM (via an all-optical magnetization
reversal)~\cite{Light_StanciuEtAl2007, Light_KirilyukEtAl2010,
  Light_HigashikawaFujitaSato2018}, by applying 
piezostrain~\cite{Piezo_PengEtAl2016, Piezo_WangEtAl2018} or by
injecting a spin-polarized current (via a spin-orbit
torque)~\cite{SOT_MironEtAl2011, SOT_LiuEtAl2012,
  SOT_GambardellaMiron2011, SOT_SinovaEtAl2015,
  SOT_ManchonEtAl2018}. This domain of research is currently under an
active exploration because of its crucial role in the implementation
of  magnetic memory and logic devices. We also note that in our
setup, the magnetic field in the 2DEG can originate from both the FM
demagnetizing field at a moderate range and from the exchange
interactions at atomic distances.

\section{Conclusions}
We have considered a Floquet triple-layer setup
of a 2DEG proximity coupled to a SC and a FM. Under resonant drive
the FM induces an oscillating Zeeman field in the 2DEG. The
out-of-plane component of the magnetic field competes with the
proximity induced SC gap and leads to the emergence of the FHeTS
hosting an effective Kramers pair of gapless helical edge modes. Moreover, the
in-plane component of the magnetic field enters into the low-energy
description corresponding to the effective Jackiw-Rebbi model as a
mass term and opens a gap in the edge mode spectrum. Change in the
sign of the mass term, which inevitably occurs at two opposite corners
of the system in a rectangular geometry, leads to the emergence
of Floquet MCSs. We argued that the proposed setup is within
experimental reach combining  available magnetic, semiconducting,
and superconducting  materials.

\section*{Acknowledgments}

We acknowledge very much the discussions with Patrick Maletinsky,
Martino Poggio, Christina Psaroudaki, Marko Ran\v{c}i\'{c}, and Flavio
Ronetti. This work was supported by the Swiss National Science
Foundation, NCCR QSIT, and the Georg H. Endress foundation. This
project received funding from the European Union's Horizon 2020
research and innovation program (ERC Starting Grant, grant agreement
No 757725).


\bibliographystyle{unsrt}


\clearpage
\widetext
\begin{center}
   \textbf{\large Supplemental Material: Floquet  Second-Order Topological Superconductor  \\Driven via Ferromagnetic Resonance}\\
  \vspace{8pt}
  Kirill Plekhanov,$^{1}$ Manisha Thakurathi,$^{1}$
  Daniel Loss,$^{1}$ and Jelena Klinovaja$^{1}$ \\ \vspace{4pt}
  $^{1}$ {\it Department of Physics, University of Basel,
    Klingelbergstrasse 82, CH-4056 Basel, Switzerland}
\end{center}

\setcounter{section}{0}
\setcounter{equation}{0}
\setcounter{figure}{0}
\setcounter{page}{1}
\makeatletter
\renewcommand{\thesection}{S\arabic{section}}
\renewcommand{\theequation}{S\arabic{equation}}
\renewcommand{\thefigure}{S\arabic{figure}}
\titleformat{\section}[hang]{\large\bfseries}{\thesection.}{5pt}{}

\section{\label{secSm:resonanceCondition}Topological criterion in the
  isotropic limit}

\begin{figure}[t]
  \centering
  \includegraphics[width=.49\columnwidth]{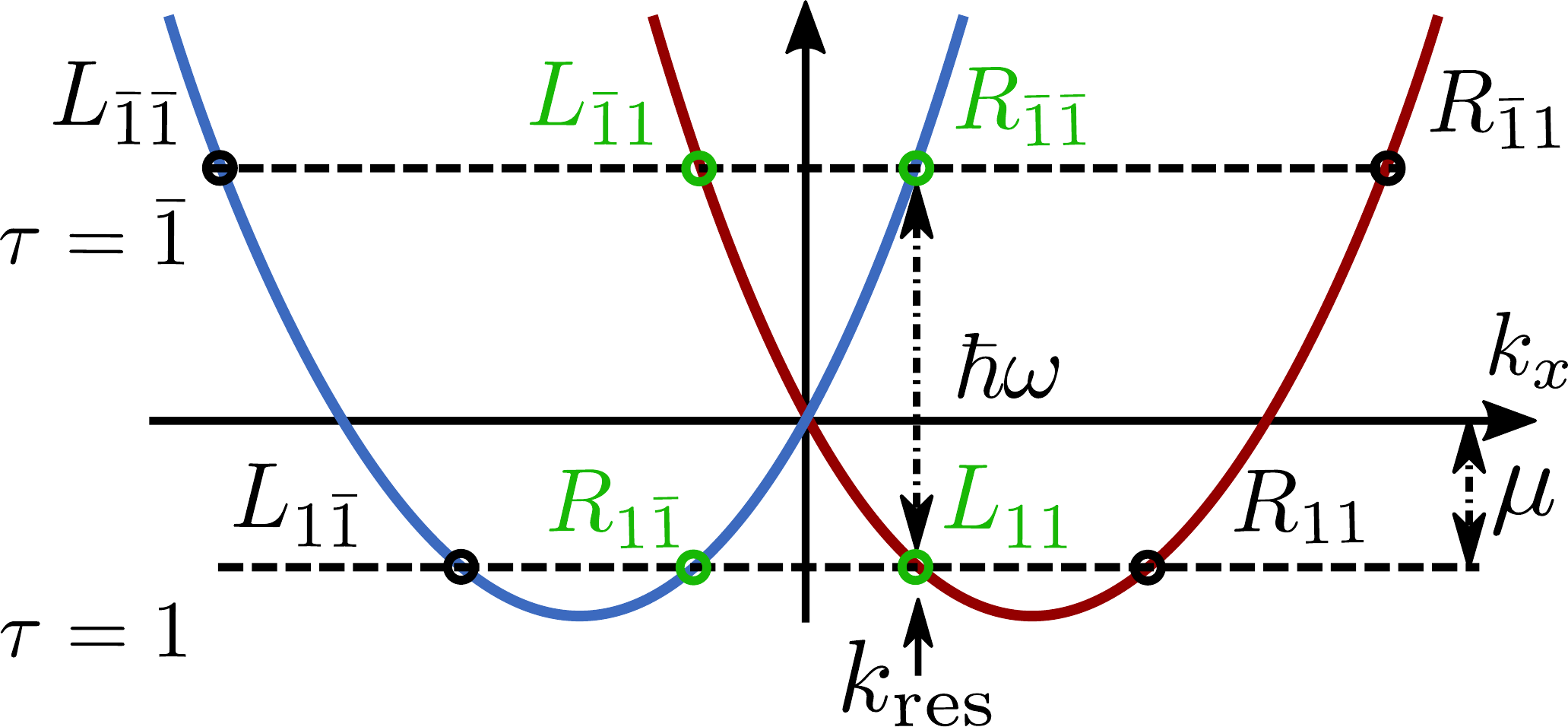}
  \caption{The band structure of the 2DEG with SOI  in the isotropic regime
    $\alpha_x = \alpha_y$ expressed as a function of $k_x$ at
    $k_y = 0$ and chemical potential $\mu$ (dashed line at bottom). The frequency $\omega$ of a periodic drive is chosen
    resonantly such that the two Floquet bands (corresponding to
    $\tau = 1, \bar{1}$) have the same smallest Fermi momentum
    $k_{\trm{res}}$ (indicated by green circles) for the two
    spin-split bands (represented by the blue and red
    colors). Operators $R_{\tau \sigma}$ and $L_{\tau \sigma}$
    correspond to the slowly varying left and right movers. The
    symbols colored in green indicate the operators involved in the
    resonant process at $k_{\trm{res}}$.}
  \label{fig:sup1-1}
\end{figure}

In the isotropic limit, $\alpha_x = \alpha_y$, when the in-plane
component of the oscillating magnetic field is zero,
$t_{\trm{Z}}^{\parallel} = 0$, the system described by the
Hamiltonian given in Eq.~(4) has a continuous
rotation symmetry in the $xy$ plane. The corresponding symmetry
operator reads 
$U^{\trm{rot}}_{J_z}(\theta) = \exp \left(-i \theta \eta_z J_z / \hbar
\right)$, with $\theta$ being the rotation angle and 
$J_z$ the $z$ component of
the orbital angular momentum operator. Thus, the energy structure depends only
on the absolute value of the momentum
$|\vec{k}| = \sqrt{k_x^2 + k_y^2}$ and can be calculated along a
particular direction in the $xy$ plane. For simplicity we perform the
calculations along the axis $k_y = 0$ [see Fig.~\ref{fig:sup1-1}].
This results in the
following Hamiltonian density
\begin{align}
  \label{eqSm:1-singleWireHam}
  \mc{H}_{\vec{k}}(k_y = 0) =
  \left( \frac{\hbar^2 k_x^2}{2m} - \mu \right)
  \eta_z
  - \alpha_x k_x
  \eta_z \sigma_y + 
  \Delta_{\trm{sc}}
  \eta_y \sigma_y
  + t_{\trm{Z}}^{\perp}
  \eta_z \tau_x \sigma_z
  + \hbar \omega
  \eta_z \left( \frac{\tau_z - \tau_0}{2} \right)
  \;.
\end{align}
Here, similar to the main text, we work in the Nambu basis
$\Psi^{\dag}_{\vec{k}}$ = ($\psi^{\dag}_{1 \uparrow \vec{k}}$,
$\psi^{\dag}_{1 \downarrow \vec{k}}$,
$\psi^{\dag}_{\bar{1} \uparrow \vec{k}}$,
$\psi^{\dag}_{\bar{1} \downarrow \vec{k}}$,
$\psi_{1 \uparrow -\vec{k}}$, $\psi_{1 \downarrow -\vec{k}}$,
$\psi_{\bar{1} \uparrow -\vec{k}}$,
$\psi_{\bar{1} \downarrow -\vec{k}}$).
We define by $\eta_i$ the Pauli
matrices (the $2 \times 2$ identity matrix for $i=0$) acting on the
particle-hole space, $\tau_i$ -- on the Floquet space, $\sigma_i$ --
on the space associated with the two spin components in the $z$
direction. The total Hamiltonian acts in the corresponding
tensor-product space and for notational simplicity we suppress the explicit writing of the tensor
product sign and the identity matrices. The system is characterized by an
effective time-reversal symmetry
$T_{\trm{eff}} = U_{T} \mc{K} = -i \tau_z \sigma_y \mc{K}$ with
$U_{T}^{\dag} \mc{H}_{\vec{k}} U_{T} = \mc{H}^{*}_{-\vec{k}}$, a
particle-hole symmetry $P = U_{P} \mc{K} = \eta_{x} \mc{K}$ with
$U_{P}^{\dag} \mc{H}_{\vec{k}} U_{P} =-\mc{H}^{*}_{-\vec{k}}$, and a
chiral symmetry $U_{C} = PT_{\trm{eff}} = -i \eta_{x} \tau_z \sigma_y$
with $U_{C}^{\dag} \mc{H}_{\vec{k}} U_{C} =-\mc{H}_{\vec{k}}$. From
this we deduce that the system belongs to the DIII symmetry class,
characterized by $\mb{Z}_2$ topological invariants, and has two
 distinct topological phases.

In the following, it will be convenient to change to the spin basis
with quantization axis along the $y$ direction, using the unitary rotation in the $yz$
plane described by the operator
$U^{\trm{rot}}_{\sigma_x} = \exp\left(-i\pi\eta_z\sigma_x / 4
\right)$. This transformation satisfies
$\tilde{\Psi}_{\vec{k}} = U^{\trm{rot}}_{\sigma_x}
\Psi_{\vec{k}}$ and
$\tilde{\mc{H}}_{\vec{k}} = U^{\trm{rot}}_{\sigma_x} \mc{H}_{\vec{k}}
{U^{\trm{rot}}_{\sigma_x}}^{\dag}$, with
\begin{equation}
  \label{eqSm:1-singleWireHam-rot}
  \tilde{\mc{H}}_{\vec{k}}(k_y = 0) =
  \left( \frac{\hbar^2 k_x^2}{2m} - \mu \right)
  \eta_z
  - \alpha_x k_x
  \sigma_z + 
  \Delta_{\trm{sc}}
  \eta_y \sigma_y
  + t_{\trm{Z}}^{\perp}
  \tau_x \sigma_y
  + \hbar \omega
  \eta_z \left( \frac{\tau_z - \tau_0}{2} \right)
  \;.
\end{equation}
In the new spin basis the Rashba SOI term acts on the spin component along
the $z$ axis and we denote by $\sigma = 1,\bar{1}$ its two possible
orientations.  In order to achieve the resonance condition, we first
fix the chemical potential $\mu$ to be smaller in absolute value than
the SOI energy,  $-E_{\trm{so}} < \mu < 0$. For instance, we restrict
the discussion to the Floquet band $\tau = 1$ only. As a result of the
Rashba SOI, which lifts the spin degeneracy, there are four Fermi
momenta
\begin{equation}
  \label{eqSm:1-1D-fermiMomenta}
  k^{F}_{1 \sigma \pm} =  \sigma k_{\trm{so}} \pm k_{\mu},
  \quad \text{with}\
  k_{\mu} = \sqrt{\frac{2 m \left( E_{\trm{so}} + \mu \right)}{\hbar^2}},\
  k_{\trm{so}} = \frac{m \alpha_x}{\hbar^2},
  \ \text{and}\
  E_{\trm{so}} = \frac{\hbar^2 k_{\trm{so}}^2}{2 m}
  \;.
\end{equation}
The resonance condition is fixed at the momentum
$k_{\trm{res}} = k^{F}_{1 1 -}$, where the spin-split band $\sigma=1$
crosses  the chemical potential  $\mu$, as shown in
Fig.~\ref{fig:sup1-1}. The resonance frequency is tuned to the
energy difference between the two spin-split bands. Since the energy
of the band $\sigma=1$ at $k_{\trm{res}}$ is zero, $\hbar \omega$ is
simply equal to the energy of the band $\sigma=\bar{1}$, which can be
written as
\begin{equation}
  \label{eqSm:1-resonanceCondition}
  \hbar \omega =
  \frac{\hbar^2 \left(k_{\trm{res}}+k_{\trm{so}} \right)^2}{2m} -
    \frac{\hbar^2 \left(k_\trm{res}-k_{\trm{so}} \right)^2}{2m} 
  = 4 E_{\trm{so}}
  \left( 1 - \sqrt{\frac{E_{\trm{so}} + \mu}{E_{\trm{so}}}} \right) \;.
\end{equation}
According to the Floquet formalism~\cite{F_Shirley1968,
  F_GoldmanDalibard2014, F_BukovEtAl2015, F_EckardtAnisimivas2015},
the second Floquet band $\tau = \bar{1}$ is shifted in energy with
respect to the first band $\tau = 1$ by an energy $-\hbar
\omega$. Hence, it crosses the chemical potential $\mu$ at four momenta
\begin{equation}
  k^{F}_{\bar{1} \sigma \pm} =
  \sigma k_{\trm{so}} \pm (2 k_{\trm{so}} - k_{\mu})
  \;.
\end{equation}
When $\omega$ is at resonance, the following identities hold true:
$k^{F}_{\bar{1} \bar{1} +} = k^{F}_{1 1 -} = k_{\trm{res}}$ and
$k^{F}_{\bar{1} 1 -} = k^{F}_{1 \bar{1} +} =-k_{\trm{res}}$.

To see the effect of $t^{\perp}_{\trm{Z}}$ and $\Delta_{\trm{sc}}$
analytically, we linearize the spectrum around the Fermi momenta
$k^{F}_{\tau \sigma \pm}$ and represent the original operators in
terms of slowly varying left and right moving
fields~\cite{RL_KlinovajaLoss2012, RL_KlinovajaLoss2015} using the
following relations
\begin{align}
  \label{eqSm:RLfields}
  \tilde{\psi}_{1 \sigma}(x)
  & =
    \tilde{R}_{1 \sigma}(x) e^{-i k^{F}_{1 \sigma +} x} +
    \tilde{L}_{1 \sigma}(x) e^{-i k^{F}_{1 \sigma -} x}
    \notag \;, \\
  \tilde{\psi}_{\bar{1} \sigma}(x)
  & =
    \tilde{R}_{\bar{1} \sigma}(x) e^{-i k^{F}_{\bar{1} \sigma +}  x} +
    \tilde{L}_{\bar{1} \sigma}(x) e^{-i k^{F}_{\bar{1} \sigma -}  x} \;.
\end{align}
In the basis associated with the slowly varying fields, the
Hamiltonian density becomes
\begin{align}
  \tilde{\mc{H}}_{\vec{k}}(k_y = 0) =
  \hbar v_F \delta k \rho_z +
  \Delta_{\trm{sc}} \rho_x \eta_y \sigma_y +
  \frac{t_{\trm{Z}}^{\perp}}{2} \tau_x
  \left( \rho_x \sigma_y - \rho_y \sigma_x \right)
  \;.
\end{align}
Here $v_F = \alpha_x / \hbar$ is the Fermi velocity assumed to be
equal for both Floquet bands,
$\delta k = k_x - k^{F}_{\tau \sigma \pm}$ is the distance from the
Fermi momenta $k^{F}_{\tau \sigma \pm}$, and the Pauli matrices $\rho_i$ act on
the space of left and right movers. The bulk spectrum of the
linearized problem is given by
\begin{align}
  E_1^2
  = (\hbar v_F \delta k)^2 + \Delta_{\trm{sc}}^2,\quad
  E_{2, \pm}^2
  = (\hbar v_F \delta k)^2 + (t_{\trm{Z}}^{\perp} \pm
  \Delta_{\trm{sc}})^2 \;.
\end{align}
Both the Floquet-Zeeman and the superconducting terms induce an
opening of the gap at the Fermi momenta $\pm k_{\trm{res}}$, which
leads to a competition between the two terms. In contrast,  the gap at the remaining Fermi
momenta is opened only by the
superconducting term. When the Floquet amplitude $t_{\trm{Z}}^{\perp}$ becomes of
the same strength as the superconducting pairing amplitude
$\Delta_{\trm{sc}}$, we observe the closing of the gap, indicating a
topological phase transition. When
$\Delta_{\trm{sc}} > t_{\trm{Z}}^{\perp}$ the system is in the
topologically trivial phase, while in the regime
$\Delta_{\trm{sc}} < t_{\trm{Z}}^{\perp}$ it hosts a Kramers pair of helical
Majorana edge modes, connected one to another via $T_{\trm{eff}}$, as
shown in Fig.~2(a) of the main text.

\section{\label{secSm:anisotropyHtsc}Boundary description of the FHeTS
  and FHOTS phases}

In order to better understand the two topological phases emerging in
our model, we study its low-energy behavior with
focus on the boundary of the system. For simplicity, we again consider
only the isotropic case $\alpha_x = \alpha_y$  here. We start with the
FHeTS phase which is characterized by the presence of helical gapless
edge modes, as shown in Fig.~2(a). We consider a
strip geometry with periodic boundary conditions along $y$ and
open boundary conditions along $x$. In such a geometry two pairs of
helical gapless modes have to be localized at the two edges at $x = 0$
and $x=L_x$. As a result of the interplay between the time-reversal symmetry $T_{\trm{eff}}$
and the particle-hole symmetry $P$, the energy of these states at
$k_y = 0$ has to be equal to zero. Hence, in order to find such states,
we look for a zero-energy solution of the real-space version of the
Hamiltonian~\eqref{eqSm:1-singleWireHam-rot}. For convenience, we keep
working in the spin basis with quantization axis along $y$ direction. We identify
the momentum $k_x$ with the spatial derivative $-i \hbar \partial_x$
and linearize the Hamiltonian around the momenta
$k^F_{\tau \sigma \pm}$ in the basis of slowly varying fields of
Eq.\eqref{eqSm:RLfields}, which leads to
\begin{align}
  \tilde{\mc{H}}(x, k_y = 0) =
  -i \hbar v_F \rho_z \partial_x +
  \Delta_{\trm{sc}} \rho_x \eta_y \sigma_y +
  \frac{t_{\trm{Z}}^{\perp}}{2} \tau_x
  \left( \rho_x \sigma_y - \rho_y \sigma_x \right)
  \;.
\end{align}
We now solve the equation
$\tilde{\mc{H}}(x, k_y = 0) \ket{\tilde{\Phi}_{\pm}} = 0$, assuming that
$\ket{\tilde{\Phi}_{\pm}}$ are two states connected to each other through the
effective time-reversal transformation
$T_{\trm{eff}} \ket{\tilde{\Phi}_{\pm}} = \pm \ket{\tilde{\Phi}_{\mp}}$. They can be
explicitly written as
$\ket{\tilde{\Phi}_{\pm}} = \sum_{x, j} \tilde{\Phi}^{j}_{\pm}(x) \ket{x, j}$ in the
orthonormal basis of spatially localized states $\ket{x, j}$ with an
orbital (Nambu) index $j$. We are particularly interested in solutions
of the form $P \ket{\tilde{\Phi}_{\pm}} = \ket{\tilde{\Phi}_{\pm}}$ associated with
self-adjoint Majorana operators. We also focus on the edge at $x=0$,
which requires imposing  vanishing boundary conditions
$\tilde{\Phi}^{j}_{\pm}(x=0) = 0$ for all $j$. We find that such solutions
exist only in the topological phase $\Delta_{\trm{sc}} < t_{\trm{Z}}^{\perp}$ and are
expressed as follows (suppressing normalization constants):
\begin{align}
  \label{eq:eqSm-majoranaKramersPair}
  &\tilde{\Phi}_{+}(x)
    = \Big(
    f_1, g_1, f_{\bar{1}}, g_{\bar{1}},
    f^*_1, g^*_1, f^*_{\bar{1}}, g^*_{\bar{1}} \Big)^{	T} ,
    \notag \\
  &\tilde{\Phi}_{-}(x)
    = \Big(
    -g_1^*,f_1^*,g_{\bar{1}}^*,-f_{\bar{1}}^*,
    -g_1,f_1,g_{\bar{1}},-f_{\bar{1}} \Big)^{T} ,
    \notag \\
  &f_{\tau}(x) = -i g_{\tau}^*(x) = e^{-i x k^{F}_{\tau \sigma +}} e^{-x /
    \xi} - e^{-i x k^{F}_{\tau \sigma -}} e^{-x / \xi_{-}} \;.
\end{align}
%
Here, $\xi = \hbar v'_F / \Delta_{\trm{sc}}$ and
$\xi_{-} = \hbar v'_F / \left( t_{\trm{Z}}^{\perp} - \Delta_{\trm{sc}}
\right)$ are two correlation lengths. Functions $f_{\tau}$ and
$g_{\tau}$ are related to each other as a result of the presence of an
additional spatial unitary symmetry in the system. This symmetry reads
as $U_{\trm{MF}} = \eta_y \sigma_x$, with
${U_{\trm{MF}} \tilde{\mc{H}}_{\vec{k}}(k_y=0) {U_{\trm{MF}}}^{\dag} =
  -\tilde{\mc{H}}_{\vec{k}}(k_y=0)}$ and
$U_{\trm{MF}} \ket{\tilde{\Phi}_{\pm}} = \ket{\tilde{\Phi}_{\pm}}$. When $k_y$ is
non-zero, it also maps the $y$ component of the momentum, i.e.,
$k_y \rightarrow - k_y$, effectively exchanging the two helical
components of the gapless edge mode.

In the second part of this section we show how the modulated in-plane
magnetic field with an amplitude $t_{\trm{Z}}^{\perp}$ couples the two
helical Majorana fermion edge modes $\ket{\tilde{\Phi}_{\pm}}$ living on the
edge $x = 0$ and leads to the emergence of the FHOTS phase hosting
MCSs. First, we write down the corresponding Hamiltonian density
in the $y$-spin basis as
\begin{align}
  \tilde{\mc{H}}_{\trm{Z}}^{\parallel} =
  t_{\trm{Z}}^{\parallel} \left( u_x
  \tau_y \sigma_x + u_y \tau_y \sigma_z \right) = 
  {U^{\trm{rot}}_{\sigma_x}}^{\dag} \mc{H}_{\trm{Z}}^{\parallel}
  U^{\trm{rot}}_{\sigma_x} = 
  {U^{\trm{rot}}_{\sigma_x}}^{\dag}
  \left[
  t_{\trm{Z}}^{\parallel} \left( u_x
  \tau_y \sigma_x + u_y \eta_z \tau_y \sigma_y \right)
  \right]
  U^{\trm{rot}}_{\sigma_x}
  \;.
\end{align}
Using the particle-hole symmetry $P$, we deduce that all the diagonal
components of $\tilde{\mc{H}}^{\parallel}_{\trm{Z}}$ in this basis of
states $\ket{\tilde{\Phi}_{\pm}}$ are  exactly zero:
\begin{align}
  \braket{\tilde{\Phi}_{\pm} | \tau_y \sigma_x | \tilde{\Phi}_{\pm}}
  & =
    \braket{\tilde{\Phi}_{\pm} | P \tau_y \sigma_x P | \tilde{\Phi}_{\pm}} = -
    \braket{\tilde{\Phi}_{\pm} | \tau_y \sigma_x | \tilde{\Phi}_{\pm}} = 0
    \notag \;, \\
  \braket{\tilde{\Phi}_{\pm} | \tau_y \sigma_z | \tilde{\Phi}_{\pm}}
  & =
    \braket{\tilde{\Phi}_{\pm} | P \tau_y \sigma_z P | \tilde{\Phi}_{\pm}} = -
    \braket{\tilde{\Phi}_{\pm} | \tau_y \sigma_z | \tilde{\Phi}_{\pm}} = 0
    \;.
\end{align}
Employing the effective time-reversal symmetry $T_{\trm{eff}}$, we
also find that all the off-diagonal terms are purely imaginary:
\begin{align}
  \braket{\tilde{\Phi}_{-} | \tau_y \sigma_x | \tilde{\Phi}_{+}}
  & =
    - \braket{\tilde{\Phi}_{+} | T_{\trm{eff}} \tau_y \sigma_x T_{\trm{eff}} | \tilde{\Phi}_{-}} =
    - \braket{\tilde{\Phi}_{+} | \tau_y \sigma_x | \tilde{\Phi}_{-}} =
    - \braket{\tilde{\Phi}_{-} | \tau_y \sigma_x | \tilde{\Phi}_{+}}^{*}
    \notag \;, \\
  \braket{\tilde{\Phi}_{-} | \tau_y \sigma_z | \tilde{\Phi}_{+}}
  & =
    - \braket{\tilde{\Phi}_{+} | T_{\trm{eff}} \tau_y \sigma_z T_{\trm{eff}} | \tilde{\Phi}_{-}} =
    - \braket{\tilde{\Phi}_{+} | \tau_y \sigma_z | \tilde{\Phi}_{-}} =
    - \braket{\tilde{\Phi}_{-} | \tau_y \sigma_z | \tilde{\Phi}_{+}}^{*}
    \;.
\end{align}
This allows us to rewrite the term
$\tilde{\mc{H}}_{\trm{Z}}^{\parallel}$ to the zeroth order in
perturbation theory as
$\tilde{\mc{H}}_{\trm{Z}}^{\parallel} =
\tilde{t}_{\trm{Z}}^{\parallel} \rho_y$, with the Pauli matrix $\rho_i$ acting in the
space of $\ket{\tilde{\Phi}_{\pm}}$ and
$\tilde{t}_{\trm{Z}}^{\parallel} = t_{\trm{Z}}^{\parallel} \mf{Im}
\braket{\tilde{\Phi}_{-} | u_x \tau_y \sigma_x + u_y \tau_y \sigma_z |
  \tilde{\Phi}_{+}}$. Moreover, using the spatial unitary symmetry
$U_{\trm{MF}}$, we find that
\begin{align}
  \braket{\tilde{\Phi}_{-} | \tau_y \sigma_z | \tilde{\Phi}_{+}}
  & = -
    \braket{\tilde{\Phi}_{-} | \tau_y \sigma_z P T_{\trm{eff}} | \tilde{\Phi}_{-}} = i
    \braket{\tilde{\Phi}_{-} | \eta_x \tau_y \tau_z \sigma_z \sigma_y | \tilde{\Phi}_{-}} = i
    \braket{\tilde{\Phi}_{-} | \eta_x \tau_x \sigma_x | \tilde{\Phi}_{-}}
    \notag \\
  & = i
    \braket{\tilde{\Phi}_{-} | U_{\trm{MF}}^{\dag}
    \eta_x \tau_x \sigma_x
    {U_{\trm{MF}}} | \tilde{\Phi}_{-}} = -i
    \braket{\tilde{\Phi}_{-} | \eta_x \tau_x \sigma_x | \tilde{\Phi}_{-}} = 0 \;.
\end{align}
Hence, the $y$ component of the magnetic field does not contribute to
the emergence of the mass term $\tilde{t}_{\trm{Z}}^{\parallel}$ on
the edge $x = 0$. As a result, the mass term can be simply expressed
using the functions $f_{\eta}$ and $g_{\eta}$ as
\begin{align}
  \tilde{t}_{\trm{Z}}^{\parallel} = 2 t_{\trm{Z}}^{\parallel} u_x \int_0^\infty dx\
  \mf{Im} \left( f^{*}_1 f^{*}_{\bar{1}} - f_1 f_{\bar{1}} - g_1
  g_{\bar{1}} + g^{*}_1 g^{*}_{\bar{1}} \right)  = -8
  t_{\trm{Z}}^{\parallel} u_x  \int_0^\infty dx\ \mf{Im} \left( f_1 f_{\bar{1}} \right)
  \;.
\end{align}
Finally, after including the kinetic term linear in momentum, the total effective
Hamiltonian density describing the $x=0$ edge of the FHeTS under the
applied magnetic field takes on the form
\begin{align}
  \mc{H}_{\trm{edge}} = \hbar v^{e}_F k_y \rho_z + \tilde{t}_{\trm{Z}}^{\parallel} \rho_y \;,
\end{align}
where $v^{e}_{F}$ is the velocity of the Majorana fermions
$\ket{\tilde{\Phi}_{\pm}}$. We recover the Hamiltonian density of the
Jackiw-Rebbi model~\cite{Jackiw1Rebbi976, JackiwSchrieffer1981}.

Starting from the above expression of $\mc{H}_{\trm{edge}}$ at
$x = 0$, it is straightforward to generalize the result to other edges
using the spatial symmetries of the problem. In particular, for the opposite
edge at $x = L_x$, one has to apply the reflection symmetry, which
simply corresponds to changing the direction of the magnetic field:
$u_x \rightarrow - u_x$. Under this transformation the mass term
$\tilde{t}_{\trm{Z}}^{\parallel}$ changes sign. The change in sign of
the mass term in the Jackiw-Rebbi model implies the presence of
zero-energy domain wall bound states~\cite{Jackiw1Rebbi976,
  JackiwSchrieffer1981}, which we identify with MCSs. Hence, an even
number of such corner states has to be present in the
system (i.e. MBSs come in pairs). Alternatively, the same result can be obtained using the
in-plane rotation symmetries. We recall that in the basis
with quantization axis along  $z$ direction the rotation operator in the $xy$ plane
reads
$U^{\trm{rot}}_{J_z}(\theta) = \exp \left(-i \theta \eta_z J_z / \hbar\right)$,
while, the spin rotation operator is given by
$U^{\trm{rot}}_{S_z}(\theta) = \exp \left(-i \theta \eta_z S_z / \hbar
\right)$, with $S_z = \hbar \sigma_z$. 
The eigenstates
$\ket{\tilde{\Phi}_{\pm}(\theta)}$ living on the edge rotated by an
angle $\theta$ with respect to the $x=0$ axis, are simply expressed as
$\ket{\tilde{\Phi}_{\pm}(\theta)} =
\tilde{U}^{\trm{rot}}_{J_z+S_z}(\theta) \ket{\tilde{\Phi}_{\pm}} =
\tilde{U}^{\trm{rot}}_{J_z}(\theta)
\tilde{U}^{\trm{rot}}_{S_z}(\theta) \ket{\tilde{\Phi}_\pm}$, where all
the rotations are expressed in the $y$-spin basis 
In particular, the opposite edge at $x = L_x$ is obtained
for $\theta = \pi$. The two perpendicular edges at $y = 0$ and $y=L_x$
in a rectangular geometry with open boundary conditions along both $x$ and $y$
are calculated using $\theta = \pm
\pi/2$. We note that in all the cases only the magnetic field
component perpendicular to the edge contributes to the mass term
$\tilde{t}^{\parallel}_{\trm{Z}}$.

\section{\label{secSm:anisotropy}Anisotropic regime}

\begin{figure}[t]
  \centering
  \includegraphics[width=.65\columnwidth]{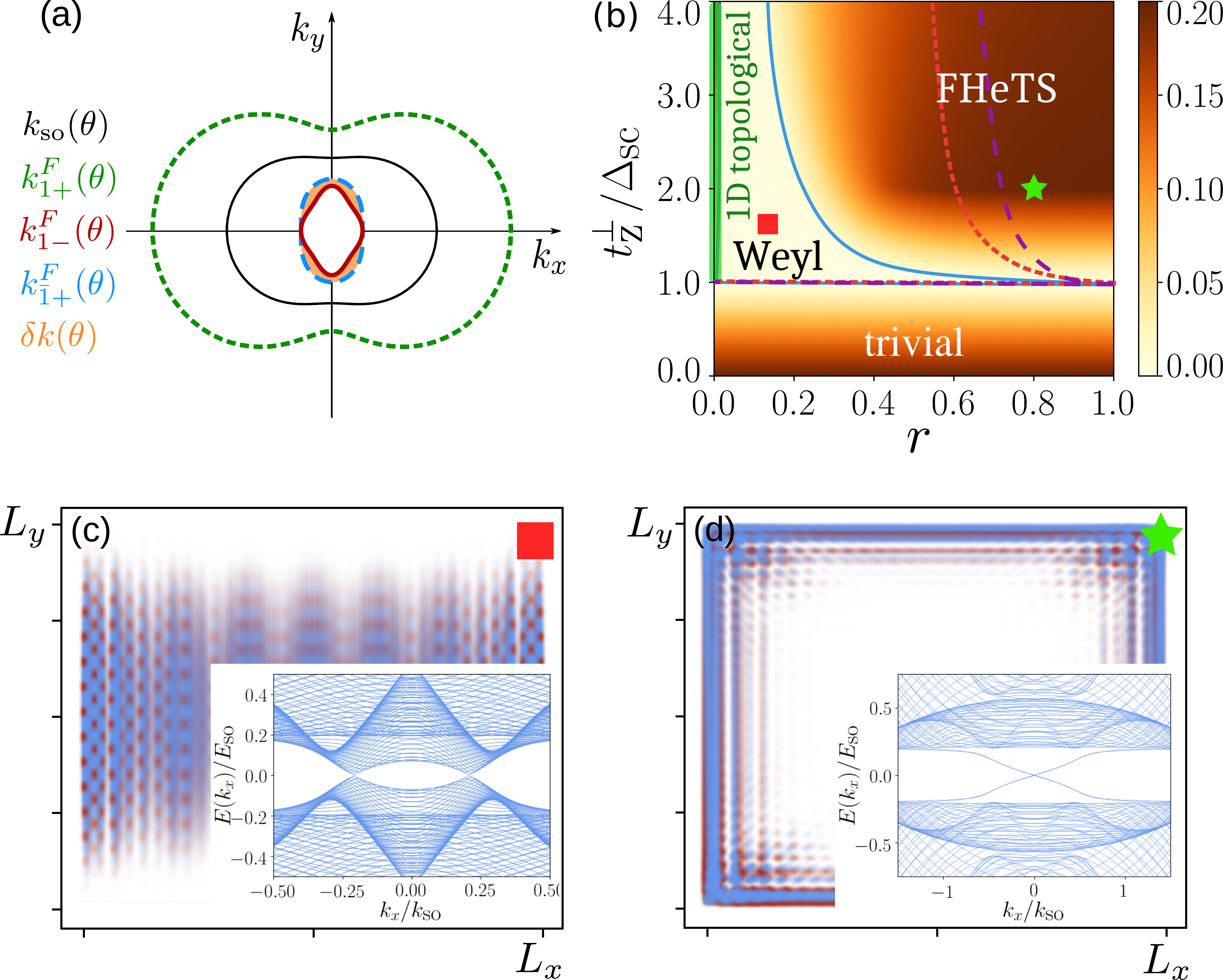}
  \caption{(a) Schematic presentation of the resonance off-set in the
    anisotropic regime  $\alpha_x \neq \alpha_y$ showing the SOI momentum
    $k_{\trm{so}}(\theta)$ (black solid line), the two Fermi surfaces
    $k^{F}_{1 \pm}(\theta)$ (red solid and green dotted lines) of the
    Floquet band $\tau = 1$ and the Fermi surface
    $k^{F}_{\bar{1} -}(\theta)$ (blue dashed line) of the Floquet band
    $\tau = \bar{1}$. The coupling between the two Floquet bands
    occurs in the neighborhood of the Fermi surfaces
    $k^{F}_{1 -}(\theta)$ and $k^{F}_{\bar{1} +}(\theta)$ but, due to
    the SOI anisotropy, a mismatch
    $\delta k = k^{F}_{\bar{1} +}(\theta) - k^{F}_{1 -}(\theta)$
    (orange filled surface) emerges. (b) Phase diagram showing the gap
    to the first excited bulk state (in units of $E_{\trm{so}}$) as a
    function of the ratios
    $r = m_x / m_y = (\alpha_y / \alpha_x)^{1/q}$ and
    $t_{\trm{Z}}^{\perp} / \Delta_{\trm{sc}}$ for $q = 1$. Different
    colors and line types of the phase boundaries correspond to
    different power laws with $q=1$ (blue solid), $q=2$ (purple
    dashed), $q=3$ (red dotted). (c)-(d) Probability density of the
    lowest energy state and the band structure $E(k_x)$ (in the inset)
    in different regions of the phase diagram. The choice of
    parameters $t^{\perp}_{\trm{F}} / \Delta_{\trm{sc}}$ and $r$ is
    indicated by the red square and the green star in (b). Remaining
    parameters in all the simulations are
    $t_{\trm{Z}}^{\parallel} = 0$,
    $\Delta_{\trm{sc}} / E_{\trm{so}} = 0.2$, $k_{\trm{so}} a = 0.2$,
    $k_{\trm{so}} L_x = k_{\trm{so}} L_y = 80$, and
    $\mu = -E_{\trm{so}} / 2$.}
  \label{fig:sup2-1}
\end{figure}

In this section we study more closely the effect of the anisotropy on
the topological phase diagram shown in Fig.~2(b). We also
consider a more general scenario when the mass $m$ is anisotropic in
the $xy$ plane with $m_x \neq m_y$. Such kind of anisotropy is more
relevant to the experimental setups of coupled Rashba wires, where the
strength of both the inter-wire SOI and the hopping term scales with
the distance between the neighboring wires.  First of all, we notice,
that when $\alpha_x \neq \alpha_y$ and (or) $m_x \neq m_y$, the Fermi
surface of the 2DEG is deformed since the rotation symmetry in the
$xy$ plane is broken. This can be seen by looking at the energy
structure, which is equal to
\begin{equation}
  E(k_x, k_y) =
  \frac{\hbar^2 k_x^2}{2 m_x} + \frac{\hbar^2 k_y^2}{2 m_y} - \mu
  \pm \sqrt{ \alpha^2_x k^2_x + \alpha^2_y k^2_y } \;.
\end{equation}
In the following it will be convenient to go to the polar coordinate
system with $k = \sqrt{k_x^2 + k_y^2}$ and
$\theta = \trm{arctan}(k_y / k_x)$. The eigenvalues
$E(k, \theta) = E(k_x, k_y)$ expressed in terms of the new coordinates
read
\begin{align}
  E(k, \theta)
  = \frac{\hbar^2 k^2}{2}
  \left(
  \frac{\cos^2 \theta}{m_x} + \frac{\sin^2 \theta}{m_y}
  \right)
  - \mu
  \pm k \sqrt{ \alpha^2_x \cos^2 \theta + \alpha^2_y \sin^2 \theta }
  = \frac{\hbar^2 k^2}{2 m(\theta)} - \mu
  \pm \alpha(\theta) k \;,
\end{align}
where we defined
$m(\theta) = m_x m_y / (m_x \sin^2 \theta + m_y \cos^2 \theta)$ and
$\alpha(\theta) = \sqrt{\alpha^2_x \cos^2 \theta + \alpha^2_y \sin^2
  \theta}$. The Fermi surface corresponds to the solutions of $k$ at a
given $\theta$ such that $E(k, \theta) = 0$. These solutions can be
explicitly written as
\begin{equation}
  k^{F}_{1 \pm}(\theta) = \left| k_{\trm{so}}(\theta) \pm k_{\mu}(\theta) \right|,
  \ \text{with}\
  k_{\mu}(\theta) = \sqrt{\frac{2 m(\theta) \left[E_{\trm{so}}(\theta) + \mu \right]}{\hbar^2}},\
  k_{\trm{so}}(\theta) = \frac{m(\theta) \alpha(\theta)}{\hbar^2},
  \ \text{and}\
  E_{\trm{so}}(\theta) = \frac{\hbar^2 k^2_{\trm{so}}(\theta)}{2 m(\theta)}
  \;.
\end{equation}
The result of the calculations is represented on
Fig.~\ref{fig:sup2-1}. In the following we will be interested only in
the solutions with the smallest amplitude $k^{F}_{1 -}(\theta)$. The
resonance condition for the frequency $\omega$ has to be fixed with
respect to a particular axis in the $xy$ plane, following the
construction procedure presented in
Section~\ref{secSm:resonanceCondition}. In this work we always fix
$\omega$ with respect to the $k_y = 0$ axis, such that
Eq.~\eqref{eqSm:1-resonanceCondition} is verified for $m = m_x$ and
$\theta = 0$. Once the choice of the axis and the frequency $\omega$
are fixed, one can calculate the solution of the equation
$E(k, \theta) = \omega$ of the form $k^{F}_{\bar{1} \pm}(\theta)$ to
deduce the Fermi surface of the second Floquet band $\tau=\bar{1}$. By
analogy to Section~\ref{secSm:resonanceCondition}, we denote the
solution with the smallest amplitude as $k^{F}_{\bar{1} +}(\theta)$,
as shown in Fig.~\ref{fig:sup2-1}(a). The resonance condition corresponds to
$k^{F}_{1-}(\theta) =k^{F}_{\bar{1}+}(\theta)$.

As a result of the anisotropy, the Fermi surfaces of the two Floquet
bands have a mismatch,
$\delta k(\theta) = k^{F}_{1-}(\theta) - k^{F}_{\bar{1}
  +}(\theta)$. This leads to the emergence of a frequency off-set
$\delta \omega \approx v_F(\theta) \delta k(\theta)$, where
$v_F(\theta) = \partial_k E(k, \theta) / \hbar$ is the Fermi velocity
evaluated at the Fermi surface. The effect of such a frequency off-set
is weaker than the one studied in Fig.~3(d) of the main
text, since it is not uniform along the entire Fermi surface and
$\delta \omega$ vanishes (by construction) for some $\theta$. In order
to study this effect more precisely, we use the numerical simulations
[see Fig.~\ref{fig:sup2-1}(b)-(d)]. We calculate the phase diagram
(the gap to the first excited bulk state) as a function of the ratios
$t_{\trm{Z}}^{\perp} / \Delta_{\trm{sc}}$, $\alpha_y / \alpha_x$, and
$m_x / m_y$. For simplicity, we assume that the two last terms scale
as a power-law: $m_x / m_y = r$, $\alpha_y / \alpha_x = r^q$. We find
that for small values of the ratio $r$ the gapped regime adiabatically
connected to the 1D topological phase transforms into a gapless regime
with several Weyl points. We also notice that for big values of $r$
the transition to the FHeTS phase is observed, similarly to
Fig.~2(b).  In order to better visualize some phases
emerging in the phase diagram, in Fig.~\ref{fig:sup2-1}(c)-(d) we
calculate the probability density of the lowest energy state and the
bulk spectrum $E(k_x, k_y)$. We see that the bulk of the FHeTS is
gapped and the boundary hosts gapless helical edge modes, while at
moderate $r$ the gap closes and four Weyl points emerge in the Weyl
phase.

\section{\label{secSm:discreteModel}Discretized model}

In the discretized version of our model, the creation (annihilation) operators
$\psi^{\dag}_{\tau \sigma m n}$ ($\psi_{\tau \sigma m n}$) of an
electron with spin component $\sigma$ along the $z$ axis, in a Floquet
band $\tau$, are defined at discrete coordinate sites $n$ and $m$. For
simplicity, we assume that the lattice constant $a$ is the same in
 $x$ and $y$ directions.
 The Hamiltonian describing the
anisotropic 2DEG, or equivalently the array of coupled Rashba wires, corresponds
to
\begin{align}
  H_{0} = \sum\limits_{mn} \sum\limits_{\tau} \Big\lbrace \Big[
  - & t_x \left(
      \psi^{\dag}_{\tau \uparrow m (n+1)} \psi_{\tau \uparrow m n} +
      \psi^{\dag}_{\tau \downarrow m (n+1)} \psi_{\tau \downarrow m n}
      \right) -
      t_y
      \left( \psi^{\dag}_{\tau \uparrow (m+1) n} \psi_{\tau \uparrow m n} +
      \psi^{\dag}_{\tau \downarrow (m+1) n} \psi_{\tau \downarrow m n}
      \right)
      \notag \\
  - & \tilde{\alpha}_x \left(
      \psi^{\dag}_{\tau \uparrow (m+1) n} \psi_{\tau \downarrow m n} -
      \psi^{\dag}_{\tau \uparrow m n} \psi_{\tau \downarrow (m+1) n}
      \right)
      + \tilde{\alpha}_y i \left(
      \psi^{\dag}_{\tau \uparrow (m+1) n} \psi_{\tau \downarrow m n} -
      \psi^{\dag}_{\tau \uparrow m n} \psi_{\tau \downarrow (m+1) n}
      \right) + \hc
      \Big]
      \notag \\
  + & \sum\limits_{\sigma}
      \left( 2 t_x + 2 t_y - \mu \right)
      \psi^{\dag}_{\tau \sigma m n}
      \psi_{\tau \sigma m n}
      \Big\rbrace
      \;.
\end{align}
Here, $t_x = \hbar^2 / (2 m_x a^2)$ and $t_y = \hbar^2 / (2 m_y
a^2)$. The spin-flip hopping amplitudes $\tilde{\alpha}_x$ and
$\tilde{\alpha}_y$ are related to the corresponding SOI strengths of the continuum model via
$\alpha_y / \tilde{\alpha}_y = \alpha_x / \tilde{\alpha}_x = 2 a$. The
proximity induced $s$-wave superconducting term is expressed as
\begin{align}
  H_{\trm{sc}} = - \frac{\Delta_{\trm{sc}}}{2} \sum\limits_{mn} \sum\limits_{\tau}
  \left(
  \psi^{\dag}_{\tau \uparrow m n} \psi^{\dag}_{\tau \downarrow m n} -
  \psi^{\dag}_{\tau \downarrow m n} \psi^{\dag}_{\tau \uparrow m n} + \hc
  \right) \;.
\end{align}
The out-of-plane component of the Floquet-Zeeman coupling is
\begin{align}
  H_{\trm{Z}}^{\perp} = t_{\trm{Z}}^{\perp} \sum\limits_{mn} \left(
  \psi^{\dag}_{1 \uparrow m n} \psi_{\bar{1} \uparrow m n} -
  \psi^{\dag}_{1 \downarrow m n} \psi_{\bar{1} \downarrow m n} + \hc
  \right) - \hbar \omega \sum\limits_{mn} \sum\limits_{\sigma}
  \psi^{\dag}_{\bar{1} \sigma m n} \psi_{\bar{1} \sigma m n} \;,
\end{align}
where the second term describes the constant energy shift of the
second Floquet band with respect to the first one. It incorporates the
$i \hbar \partial_t$-term which is present in the expression of the
quasi-energy operator written in the basis of $T$-periodic states. We
finally express the in-plane component of the Floquet-Zeeman term as
\begin{align}
  H_{\trm{Z}}^{\parallel} = t_{\trm{Z}}^{\parallel} \sum\limits_{mn}
  \sum\limits_{\tau\sigma\sigma'} \psi^{\dag}_{\tau \sigma m n}
  \left[ \vec{u}_{\parallel} \cdot \vec{\sigma} \right]_{\sigma \sigma'}
  \psi_{\bar{\tau} \sigma' m n}
  \;.
\end{align}
In the presence of  translation symmetry, the total Hamiltonian $H =
H_{0} + H_{\trm{sc}} + H_{\trm{Z}}^{\perp} +
H_{\trm{Z}}^{\parallel}$ can be diagonalized in  momentum space as
$H = \sum_{\vec{k}} \Psi^{\dag}_{\vec{k}} \mc{H}_{\vec{k}}
\Psi_{\vec{k}} / 2$, leading to
\begin{align}
  \mc{H}_{\vec{k}} = & \Big( 2 t_x \left[ 1 - \cos(k_x a) \right] + 2
  t_y \left[ 1 - \cos(k_y a) \right] - \mu \Big) \eta_z - 2
  \tilde{\alpha}_x \sin(k_x a) \eta_z \sigma_y + 2 \tilde{\alpha}_y
  \sin(k_y a) \sigma_x \notag \\ + & \Delta_{\trm{sc}} \eta_y \sigma_y
  + t_{\trm{Z}}^{\perp} \eta_z \tau_x \sigma_z +
  t_{\trm{Z}}^{\parallel} \left( u_x \tau_y \sigma_x + u_y \eta_z
  \tau_y \sigma_y \right) + \hbar \omega \eta_z \left( \frac{\tau_z -
    \tau_0}{2} \right) \;.
\end{align}
The explicit  representation of the Hamiltonian as $8\times 8 $ matrix reads
{\footnotesize
\begin{equation}
  \mc{H}_{\vec{k}} =
  \left(
  \begin{array}{cccccccc}
    \lambda_{\vec{ k}} & \alpha_{ k_y}+i\alpha_{ k_x} & t_{\trm{Z}}^{\perp} &
    t_{\trm{Z}}^{\parallel} \left(-u_y - i u_x \right) & 0 &-\Delta_{\trm{sc}} & 0 & 0 \\
    \alpha_{ k_y}-i\alpha_{ k_x} & \lambda_{\vec{ k}} &
    t_{\trm{Z}}^{\parallel} \left( u_y - i u_x \right) &-t_{\trm{Z}}^{\perp} & \Delta_{\trm{sc}} & 0 & 0 & 0 \\
    t_{\trm{Z}}^{\perp} & t_{\trm{Z}}^{\parallel} \left( u_y + i u_x \right) & \lambda_{\vec{ k}}-\hbar\omega
    & \alpha_{ k_y}+i\alpha_{ k_x} & 0 & 0 & 0 &-\Delta_{\trm{sc}} \\
    t_{\trm{Z}}^{\parallel} \left(-u_y + i u_x \right) &-t_{\trm{Z}}^{\perp} & \alpha_{ k_y}-i\alpha_{ k_x}
    & \lambda_{\vec{ k}}-\hbar\omega & 0 & 0 & \Delta_{\trm{sc}} & 0 \\
    0 & \Delta_{\trm{sc}} & 0 & 0 &-\lambda_{\vec{k}} &
   \alpha_{k_y}-i\alpha_{k_x} &-t_{\trm{Z}\perp} & t_{\trm{Z}\parallel} \left( u_y-iu_x \right) \\
   -\Delta_{\trm{sc}} & 0 & 0 & 0 & \alpha_{k_y}+i\alpha_{k_x}
   &-\lambda_{\vec{k}} & t_{\trm{Z}}^{\parallel} \left(-u_y - i u_x \right) & t_{\trm{Z}}^{\perp} \\
   0 & 0 & 0 & \Delta_{\trm{sc}} & -t_{\trm{Z}}^{\perp} & t_{\trm{Z}}^{\parallel} \left(-u_y + i u_x \right)
   &-\lambda_{\vec{k}}+\hbar\omega & \alpha_{k_y}-i\alpha_{k_x} \\
   0 & 0 &-\Delta_{\trm{sc}} & 0 & t_{\trm{Z}}^{\parallel} \left( u_y + i u_x \right)
   & t_{\trm{Z}}^{\perp} & \alpha_{k_y}+i\alpha_{k_x} &-\lambda_{\vec{k}}+\hbar\omega
  \end{array} \right),
\end{equation}}

\hspace{-15pt}
where we introduced the notations
$\lambda_{\vec{k}} = t_x \left[ 1 - \cos(k_x a) \right] + t_y \left[ 1 -
  \cos(k_y a) \right] - \mu / 2$,
${\alpha_{k_x} = \tilde{\alpha}_x \sin(k_x a)}$, and
${\alpha_{k_y} = \tilde{\alpha}_y \sin(k_y a)}$.

\section{\label{secSm:oommf} Micromagnetic simulations}

As shown in the main text, to reach the topological phase transition
one needs to generate a magnetic field $\vec{B}(t)$ with a magnitude
of the order of $0.1$ T oscillating in the GHz frequency
range. Moreover, the oscillating component of $\vec{B}(t)$ should be
greater than the static one. One possible solution to generate such a
magnetic field consists in placing the 2DEG layer in proximity to a FM
slab. In this section we provide details of this construction.

At equilibrium, the magnetization $\vec{M}$ of the FM is aligned
with the easy axis of the FM. For simplicity, we assume that the easy axis
lies in the $xy$ plane, parallel to the bottom (and top) surface of the FM slab. We
consider a time-dependent protocol, where the system at an initial
time is subjected to a static external magnetic field $\vec{H}_0$,
which makes an angle with the magnetization $\vec{M}$ but also lies in
the $xy$ plane. Moreover, an additional small time-dependent magnetic
field $\vec{h}(t)$ with $|\vec{h}(t)| \ll |\vec{H}_0|$ is applied
perpendicularly to $\vec{H}_0$ such that the total applied external
magnetic field is $\vec{H}(t) = \vec{H}_0 + \vec{h}(t)$. As a result,
the magnetization $\vec{M}(t)$ will start to precess, described by the
Landau-Lifshitz-Gilbert (LLG) equation~\cite{LandauLifshitz1935,
  Gilbert1955}
\begin{equation}
  \frac{\trm{d} \vec{M}(t)}{\trm{d} t} =
  - \gamma \vec{M}(t) \times \vec{B}(t)
  - \alpha \frac{\gamma}{M_{\trm{s}}} \vec{M}(t) \times
  \left[ \vec{M}(t) \times \vec{B}(t) \right] \;.
\end{equation}
Here, $\gamma$ is the gyromagnetic ratio, $\alpha$  the dimensionless
damping factor, and $M_{\trm{s}}$  the saturation magnetization.
The total magnetic field $\vec{B}(t)=\vec{H}(t)+\vec{D}(t)$ comprises the applied
static field $\vec{H}_0$, the small time-dependent field $\vec{h}(t)$,
and the demagnetizing field $\vec{D}(t)$ generated by the FM. We
note that the demagnetizing field has in general a complicated
structure that depends on the shape of the FM as well as the
crystalline anisotropy. The LLG equation can be solved under
simplified conditions, by neglecting the small field $\vec{h}(t)$ and
assuming that the demagnetizing field is constant:
$\vec{D}(t) = \vec{D}$. Such a solution describes a damped precession
of the magnetization with the frequency determined by the Kittel
formula~\cite{Kittel1948, Kittel1951}
\begin{equation}
  \omega = \gamma \sqrt{|\vec{H}_0| | \vec{H}_0  + \vec{D} |} \;.
\end{equation}
Finally, if the small field $\vec{h}(t)$ is resonant at the frequency
$\omega$, the FM absorbs energy, establishing a steady-state
with a fixed precession angle of the magnetization.

\begin{figure}[t]
  \centering
  \includegraphics[width=.99\columnwidth]{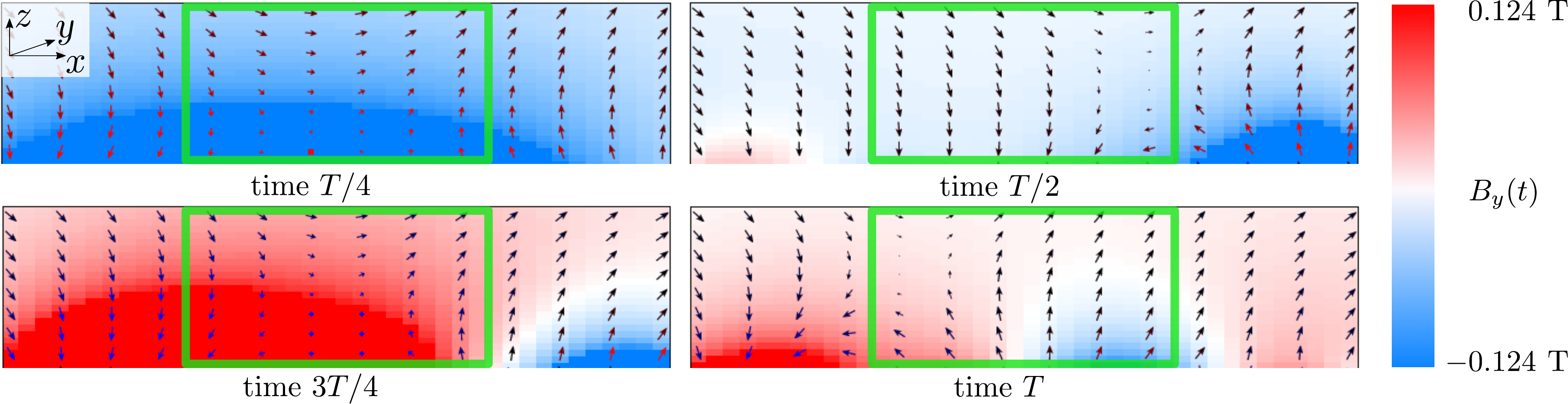}
  \caption{OOMMF simulations of the total effective field $\vec{B}(t)$
    in a $50$ nm height pocket above the FM at four different  moments of time
     in the interval $[0, T]$. Arrows represent the $x$ and $z$
    components of  $\vec{B}(t)$, while the intensity plot represents its
    $y$ component. In the region inside the green box the static
    component of $\vec{B}(t)$ is much smaller than the oscillating
    part, and the effective field makes a full rotation in the $yz$
    plane, corresponding to the optimal arrangement for our Floquet
    setup.
    In the region outside the green box the left (right) boundary of the
    FM is too close and no sizable precession of $\vec{B}(t)$ can
    develop there. The amplitude of the magnetic field is shown on the scale
    $[0, 0.124]$ T. Parameters
    of the simulation are
    $L_x \times L_y \times L_z = 200 \times 200 \times 100$ nm and
    $|\vec{H}_0| = 0.124$ T.}
  \label{fig:sup3-1}
\end{figure}

The protocol described above allows one to generate a substantial large
oscillating magnetic field through the demagnetizing field of the
FM. However, generically the static component of such a magnetic
field will dominate over the dynamic one, which is not desired for
our purpose. This problem can be removed by adjusting the geometry of the
setup (the size of the FM) and the strength of the applied static
field $\vec{H}_0$, so that the static component of the demagnetizing
field $\vec{D}(t)$ cancels $\vec{H}_0$ exactly. In Fig.~\ref{fig:sup3-1} we
show the result of a calculation, based on the numerical solution of the LLG
equation using the finite difference micromagnetic solver
OOMMF~\cite{oommf}. We find that for a FM slab of the size
$L_x \times L_y \times L_z = 200 \times 200 \times 100$ nm there is a
wide region of the space distanced from the left and right boundaries of the FM, where the
total effective field $\vec{B}(t)$ oscillates at the frequency
$\omega = 2 \pi / T = 63$ GHz and makes a full $360^{\circ}$ rotation
in the $yz$ plane (which is optimal for our purpose). The initial parameters of the simulation are chosen
such that the static field $\vec{H}_0$ with $|\vec{H}_0| = 0.124$ T
is directed along the $x$ axis. At the initial time the
magnetization $\vec{M}$ makes an angle of
$11.3^{\circ}$ with the vector $\vec{H}_0$. The amplitude of the
resulting oscillating field $\vec{B}(t)$ is of the order of $0.1$ T.


\bibliographystyle{unsrt}

\begin{thebibliography}{1}
\bibitem{fot1} The initial presence of the effective symmerty $T_{\trm{eff}}$  is not crucial for our setup, as this symmetry is broken by the in-plane field anyway, thus,  our setup is stable against disorder and does not rely on the presence of particular spatial symmetries.
\end{thebibliography}


\end{document}